\documentclass[12pt]{article}
\pdfoutput=1
\usepackage{yfonts}
\usepackage{color}
\usepackage{mhchem}
\usepackage{xcolor}
\usepackage{cite}
\usepackage{hyperref}
\hypersetup{colorlinks=true,linkcolor=red,anchorcolor=black,citecolor=green}
\usepackage[toc,page]{appendix}
\usepackage{amsfonts}
\usepackage{bbold}
\usepackage{textcomp}
\usepackage[DIV13]{typearea}
\usepackage{amsmath, amsthm, amssymb, mathtools,empheq,latexsym,dsfont}
\usepackage{bbm}
\usepackage{slashed, simplewick}
\usepackage[utf8]{inputenc}
\usepackage{graphicx,placeins}
\usepackage{makeidx}
\usepackage[font=small,labelfont=bf]{caption}
\usepackage{nicefrac}
\usepackage{subfigure}
\usepackage{array, bigdelim,multirow,multicol}
\usepackage[integrals]{wasysym}
\usepackage{fancybox}
\usepackage{bm}
\usepackage{float}
\usepackage{rotating}
\usepackage{colortbl}
\usepackage{booktabs}
\usepackage[top=2cm,textwidth=16.6cm,textheight=22.75cm]{geometry}
\usepackage{doi}
\usepackage{fancybox}
\usepackage[compat=1.0.0]{tikz-feynman}
\usepackage{tikz}
\usepackage{empheq}
\usepackage{longtable}
\allowdisplaybreaks
\newcommand{\ignore}[1]{}

\usepackage{braket}
\usepackage{makecell}
\usepackage{diagbox}
\usepackage{enumitem}  

\graphicspath{{immagini/}}

\definecolor{Gray}{gray}{0.92}

\DeclareMathAlphabet{\mathscr}{OT1}{pzc}{m}{it}

\newcommand{\be}{\begin{equation}}
\newcommand{\ee}{\end{equation}}
\newcommand{\bea}{\begin{eqnarray}}
\newcommand{\eea}{\end{eqnarray}}

\makeatletter
\@addtoreset{equation}{section}
\makeatother

\begin{document}

\unitlength = 1mm
\setlength{\extrarowheight}{0.2 cm}
\thispagestyle{empty}
\bigskip
\vskip 1cm

\title{\Large \bf Two-zero textures of the Majorana neutrino mass matrix from $\mathbb{Z}_3$ gauging of $\mathbb{Z}_N$ non-invertible symmetry \\[2mm] }

\date{}

\author{Bu-Yao Qu$^{a}$\footnote{E-mail: {\tt
qubuyao@mail.ustc.edu.cn}},  \
Zheng Jiang$^{a}$\footnote{E-mail: {\tt
jzdyx@mail.ustc.edu.cn}},\
Gui-Jun Ding$^{a,b}$\footnote{E-mail: {\tt
dinggj@ustc.edu.cn}}
\\*[20pt]
\centerline{
\begin{minipage}{\linewidth}
\begin{center}
$^a${\it \small Department of Modern Physics,  and Anhui Center for fundamental sciences in theoretical physics,\\
University of Science and Technology of China, Hefei, Anhui 230026, China}\\[2mm]
$^b${\it \small College of Physics, Guizhou University, Guiyang 550025, China}
\end{center}
\end{minipage}}
\\[10mm]}
\maketitle
\thispagestyle{empty}

\centerline{\large\bf Abstract}

\begin{quote}
\indent

Texture-zero ansatze offer an economical description of neutrino masses, with current data allowing only seven inequivalent two-zero Majorana textures in the charged-lepton mass basis. We investigate how such textures can arise from non-invertible symmetries realized through $\mathbb{Z}_3$ gauging of $\mathbb{Z}_N$. In contrast to $\mathbb{Z}_2$ gauging, which necessarily induces diagonal neutrino mass terms via the Weinberg operator, $\mathbb{Z}_3$ gauging admits complex representations and allows a richer class of neutrino mass textures.  
If the light neutrino mass is described by the Weinberg operator, we find that the textures $\mathbf{A}_{1,2}$, $\mathbf{B}_{3,4}$, and $\mathbf{C}$ can be realized from the $\mathbb{Z}_{3}$ gauging of $\mathbb{Z}_{13}$ symmetry, while all the seven phenomenologically viable two-zero textures can emerge from  $\mathbb{Z}_{3}$ gauging of $\mathbb{Z}_{19}$ symmetry without requiring supersymmetry. When the neutrino mass is generated by the type-I seesaw mechanism, the structure of the non-invertible symmetry is more restrictive, yielding only texture $\mathbf{C}$ for $N\neq7$. 
These results demonstrate the strong predictive power of non-invertible symmetries for neutrino mass textures. Furthermore, the more general $\mathbb{Z}_{n}$ gauging of the $\mathbb{Z}_{N}$ symmetry with $n>3$ is analyzed, which results in novel fusion rules.

\end{quote}

\clearpage

\section{Introduction}

Origin of fermion mass and flavor mixing is a longstanding open question of particle physics, and it is known as flavor puzzle~\cite{Xing:2020ijf,Feruglio:2019ybq,Ding:2023htn,Ding:2024ozt}.  Several theoretical approaches have been developed to address the flavor puzzle. One way to reduce the number of free parameters in fermion mass matrices is to assume that some of their entries vanish, leading to the so-called texture-zero models~\cite{Fritzsch:1977za,Weinberg:1977hb,Wilczek:1977uh}. See Refs.~\cite{Fritzsch:1999ee,Gupta:2012fsl} for comprehensive review on the approach of texture-zero.  It has been shown that Majorana neutrino mass matrices with more than two texture zeros are incompatible with current neutrino oscillation data~\cite{Frampton:2002yf,Xing:2002ta}. By contrast, textures with only one zero exhibit limited predictive power. In the charged lepton diagonal basis, there are only seven patterns of two-zero texture light neutrino mass matrix compatible with current experimental data at the $3\sigma$ level, they are denoted as $\textbf{A}_{1,2}$, $\textbf{B}_{1,2,3,4}$ and $\textbf{C}$ with the following texture:
\begin{eqnarray}
\nonumber &&\textbf{A}_1~:~\begin{pmatrix}
0&0&\times \\
0&\times &\times \\
\times &\times &\times 
\end{pmatrix}\,,~~~\textbf{A}_2~:~\begin{pmatrix}
0&\times&0 \\
\times &\times &\times \\
0&\times &\times
\end{pmatrix}\,,~~~\textbf{B}_1~:~\begin{pmatrix}
\times&\times& 0 \\
\times & 0&\times \\
0 &\times &\times 
\end{pmatrix}\,,~~~\textbf{B}_2~:~\begin{pmatrix}
\times&0& \times \\
0 &\times &\times \\
\times &\times &0 
\end{pmatrix}\,,\\
&&\textbf{B}_3~:~\begin{pmatrix}
\times& 0& \times \\
0 & 0&\times \\
\times &\times &\times 
\end{pmatrix}\,,~~~
\textbf{B}_4~:~\begin{pmatrix}
\times&\times& 0 \\
\times & \times &\times \\
0 &\times &0 
\end{pmatrix}\,,~~~\textbf{C}~:~\begin{pmatrix}
\times& \times& \times \\
\times & 0&\times \\
\times &\times &0
\end{pmatrix}\,, \label{eq:ABC-2zero}
\end{eqnarray}
where $\times$ denotes non-vanishing entry.
It is remarkable that these textures not only successfully account for the observed lepton mixing pattern but also provide nontrivial predictions for the absolute neutrino mass scale and the CP-violating phases.

The texture-zero of neutrino mass matrix is phenomenologically appealing, they yield strong and testable predictions. In theoretical constructions, the texture zero may reflect some underlying selection rules, and it can be generated from the Abelian~\cite{Grimus:2004hf,GonzalezFelipe:2014zjk} or non-Abelian discrete flavor symmetries. In certain non-linear realizations of flavor symmetry, such as modular symmetry, the nonzero entries of the mass matrices can be correlated, thereby enhancing the predictive power of texture-zero models~\cite{Zhang:2019ngf,Lu:2019vgm,Kikuchi:2022svo,Ding:2022aoe}. Over the last decades, the concept of global symmetry has been significantly generalized, motivated by the seminal work~\cite{Gaiotto:2014kfa}. Among these developments, various classes of non-invertible symmetries have attracted considerable attention~\cite{Schafer-Nameki:2023jdn,Bhardwaj:2023kri,Shao:2023gho}. Non-invertible symmetries do not possess a group structure, making them promising candidates for explaining texture zeros of neutrino mass matrices and Yukawa couplings~\cite{Kobayashi:2024cvp,Kobayashi:2025znw,Kobayashi:2025ldi,Jiang:2025psz}. Moreover, realistic quark textures realized through non-invertible symmetries can provide an axion-less solution to the strong CP problem~\cite{Liang:2025dkm,Kobayashi:2025rpx}.

Previous studies have shown that $\mathbb{Z}_2$ gauging of $\mathbb{Z}_N$ symmetry can be employed to realize phenomenologically interesting Yukawa textures. However, such constructions are unable to generate the viable two-zero textures of neutrino mass matrix shown in Eq.~\eqref{eq:ABC-2zero}, since the diagonal entries of the neutrino mass matrix can not be forbidden by the $\mathbb{Z}_2$ gauging of $\mathbb{Z}_N$ symmetry alone if neutrino mass is generated by the Weinberg operator.   The $\mathbb{Z}_3$ gauging of $\mathbb{Z}_N$ symmetry, on the other hand, possesses properties that differ substantially from those of $\mathbb{Z}_2$ gauging~\cite{Dong:2025jra,Jangid:2025krp,Kobayashi:2025wty,Okada:2025adm}.   
The purpose of this paper is to employ the selection rules of the $\mathbb{Z}_3$ gauging of $\mathbb{Z}_N$ symmetry to realize the two-zero textures of Majorana neutrino mass matrix, where fields are labeled by elements of the underlying algebra. Notably, for suitable values of $N$, all texture-zero patterns can be generated from the Weinberg operator, whereas only the texture $\mathbf{C}$ arises when the light neutrino masses are generated via the type-I seesaw mechanism.

The remainder of this paper is organized as follows. In section~\ref{sec:Z3_gauging_ZN}, we review the formalism of $\mathbb{Z}_3$ gauging of $\mathbb{Z}_N$ symmetry. Section~\ref{sec:neu_mass} is devoted to reproducing the allowed two-texture-zero neutrino mass matrices from $\mathbb{Z}_3$ gauging of $\mathbb{Z}_N$ for $N = 13$ and $19$, together with an analysis of the corresponding charged-lepton mass matrix. We also examine the neutrino mass textures generated within the type-I seesaw framework. In section~\ref{sec:Zn_gauging_ZN}, we extend the $\mathbb{Z}_3$ gauging of $\mathbb{Z}_N$ to the more general $\mathbb{Z}_n$ gauging of $\mathbb{Z}_N$, new fusion algebras are obtained. Finally, our main conclusions are summarized in section ~\ref{sec:conclusion}. The phenomenological implications of two-zero texture for neutrino oscillation parameters are discussed in appendix~\ref{sec:phenomenon}.
Likewise, the $\mathbb{Z}_3$ gauging of $\mathbb{Z}_{7}$ and $\mathbb{Z}_9$ symmetries constrains the possible structures of the neutrino mass matrix, as summarized in Appendix~\ref{sec:Z3_gauging_of_Z7_Z9}. Furthermore, the more general schemes of $\mathbb{Z}_2$ gauging of $\mathbb{Z}_N$ symmetry are discussed in Appendix~\ref{sec:Z2_gauging_of_ZN}.

\section{$\mathbb{Z}_3$ gauging of $\mathbb{Z}_N$ symmetries \label{sec:Z3_gauging_ZN}}

Gauging invertible symmetries is an important way to construct non-invertible symmetries. Let us consider a theory with finite symmetry group $T_N \cong \mathbb{Z}_N \rtimes \mathbb{Z}_3$~\cite{Ishimori:2010au}. The group $T_7$~\cite{Luhn:2007sy,Hagedorn:2008bc,King:2009ap} and $T_{13}$~\cite{Ding:2011qt,Hartmann:2011pq} as the flavor symmetry have been studied in these literature. The $\mathbb{Z}_N$ subgroup of $T_N$ is generated by $a$ with $a^N = e$, and the $\mathbb{Z}_3$ subgroup generated by $b$ factor acts as a nontrivial automorphism of $\mathbb{Z}_N$ as follow, 
\begin{eqnarray}\label{eq:Z3_automorphism}
b^{-1}a b = a^m \,,~~~ b^3 = e\,,
\end{eqnarray} 
where $e$ refers to the identity element, $m$ is an integer whose value depends on $N$. Applying the automorphism $b$ successively, one finds
\begin{eqnarray}
a = b^{-3} a b^3 = b^{-2} a^m b^2 = b (b^{-1} a b)^m b = b^{-1} a^{m^2} b = ( b^{-1} a b )^{m^2} = a^{m^3}  \,.
\end{eqnarray}
This implies 
\begin{eqnarray}\label{eq:consist_n3}
m^3 - 1 = (m-1) (m^2 + m + 1) \equiv 0 ~ ({\rm mod}~ N) \,.
\end{eqnarray}
For any given solution $m=m_0$ to Eq.~\eqref{eq:consist_n3}, $m=m_0^2~({\rm mod}~N)$ also satisfies Eq.~\eqref{eq:consist_n3}.
We see that one of the solution is $m=1~({\rm mod}~N)$, but this is trivial since $a$ would commute with $b$ and the resulting group is isomorphic to $\mathbb{Z}_N \times \mathbb{Z}_3$ rather than $T_N$. As a consequence, we are concerned with the solution of $m\neq 1$ in the following. Firstly we study the case that $N$ is a prime number or its power except $3$, i.e. $N=p^\alpha$, where $p\neq3$ is a prime number and $\alpha$ is a positive integer. Then $m-1$ cannot be divisible by $p$. Hence, $m$ must satisfy
\begin{eqnarray}
m^2 + m + 1 \equiv 0 ~ ({\rm mod}~ N) \,.
\end{eqnarray}
The possible combinations of $(N,m)$ are
\begin{eqnarray}
(N, m) = (7,2),  (13,3),  (19,7) \cdots  \,.
\end{eqnarray}
It can be shown that such $N$ must be power of a prime number $p$ of the form $p=3k+1$ where $k$ is a positive integer.

If $N$ is the power of $3$ with $N = 3^\alpha$ with $\alpha\geq 2$, then $m-1$ should be divisible by $3^{\alpha-1}$, since $m^2 + m + 1$ is divisible at most by $3$~\footnote{If $m^2 + m + 1$ is not divisble by $3$, then the unique solution will be $m=1~ ({\rm mod}~ N)$, which is the trivial solution.}. As a consequence, we have $m = 3^{\alpha-1}+1$ or $m= 2 \times 3^{\alpha-1}+1$. 
If $N$ is not the power of a prime number, that is, if $N$ contains different prime factors, then the corresponding values of $N$ and $m$ can be obtained from the above results~\footnote{Generally $N$ can be factorized as $N = p_1^{\alpha_1} \cdots p_s^{\alpha_s}$ where $p_i$ are prime numbers, the equation $m^3 -1 \equiv 0 ~ ({\rm mod}~ N)$ can be decomposed into the system
\begin{eqnarray}
m^3 - 1 \equiv 0 ~ ({\rm mod}~ p_1^{\alpha_1}) \,, \cdots \,, m^3 - 1 \equiv 0 ~ ({\rm mod}~ p_s^{\alpha_s})\,.
\end{eqnarray}
In order to obtain a value of $m\neq 1 ~({\rm mod}~N)$, 
at least one equation in the above system must have non-trivial solution.
Consequently, one must have either $9|N$ or $p|N$ for some prime number $p$ of the form $p=3k+1$. 
 }.
Here we list some possible values of $N$ and $m$ in table~\ref{tab:N-m}, where the trivial solution of $m=1$ is dropped. Note that $m$ can have more than two distinct values if  $N$ is large. For instance, we have $m= 4\,, 16\,, 22\,, 25\,, 37\,, 43\,, 46\,, 58$ for $N=63$.

\begin{table}[t!]
\centering
\begin{tabular}{|c||c|c|c|c|c|c|c|}
\hline \hline
$N$ & $7$ & $9$ & $13$ & $14$ & $18$ & $19$ & $21$  \\ \hline 
$m$ & $2,4$ & $4,7$ & $3,9$ & $9, 11$ & $7,13$ & $7, 11$ & $4,16$  \\ \hline \hline
\end{tabular}
\caption{\label{tab:N-m}Some small values of $N$ and $m$ which satisfy the equation $m^3-1 = 0 ~ ({\rm mod}~ N) $.  }
\end{table}

Gauging $\mathbb{Z}_3$ produces the theory with symmetry category leading to non-invertible structures. 
The simple objects are the $\mathbb{Z}_3$ orbits of $\mathbb{Z}_N$. Intuitively, the elements within each orbit are glued together.
For a generic element $a^k$, we find 
\begin{eqnarray}
b a^k b^{-1} =(bab^{-1})^k= a^{m^2 k} \,,~~~b^2 a^k b^{-2} = b^{-1} a^k b = (b^{-1} a b)^k =(a^m)^k=a^{mk}\,,
\end{eqnarray}
where we have used the identity $bab^{-1}=b^{-2}ab^{2}=b^{-1}\left(b^{-1}ab\right)b=b^{-1}a^mb=\left(b^{-1}ab\right)^m=(a^m)^m=a^{m^2}$.
Then, we can obtain the following equivalence classes by $\mathbb{Z}_3$ orbit:
\begin{eqnarray}
\label{eq:EQ-classes}C^{(k)} = \{ a^k, a^{km}, a^{k m^2} \} \equiv [k] \,,
\end{eqnarray}
where $k$ and $m$ are generic integers. The class $[0]$  comprises only the identity, i.e. $[0] = \{ e \} $. These equivalence classes are parts of the conjugacy classes of $T_N$. The multiplication rules between different equivalence classes are given by
\begin{eqnarray}
\nonumber [k] \times [l] &=& \{ a^k, a^{km}, a^{k m^2} \} \times \{ a^l, a^{lm}, a^{l m^2} \} \\
\nonumber &=& \{ a^{k+l}, a^{(k+l)m}, a^{(k+l)m^2} \} + \{ a^{k+lm}, a^{(k+lm)m}, a^{(k+lm)m^2} \} + \{ a^{k+lm^2}, a^{(k+lm^2)m}, a^{(k+lm^2)m^2} \}  \\
&=& [k+l] + [k+lm] + [k+lm^2]\,, \label{eq:fussion-rules-gener}
\end{eqnarray}
which implies
\begin{equation}
[k]\times [l] =[l] \times [k]\,. 
\end{equation}
From Eq.~\eqref{eq:fussion-rules-gener}, we can see that there exist a unique inverse class $[-k]=\{a^{-k}, a^{-km}, a^{-km^2}\}$ for each equivalence class $[k]$,  such that we have $[0]\subset[k]\times [-k]$.
The fusion algebra reflects the orbit structure for the $\mathbb{Z}_3$ action on $\mathbb{Z}_N$. The full conjugacy classes of $T_N$ include the following two additional classes,
\begin{eqnarray}
C_N^{(1)} = \{ b, ba, \cdots , ba^{N-1} \}~~~ \text{ and }~~~ C_N^{(2)} = \{ b^2, b^2a, \cdots , b^2a^{N-1} \} \,.
\end{eqnarray}
The non-invertible $\mathbb{Z}_3$ gauging of $\mathbb{Z}_N$ symmetry can be realized in higher dimensional field theory in a way similar to that of $\mathbb{Z}_2$ gauging of $\mathbb{Z}_N$ symmetry. We start with a field $\varphi_k$ with definite $\mathbb{Z}_N$ charge $\varphi_k \rightarrow \zeta^{k} \varphi_k$, $\zeta\equiv e^{2\pi i/N}$. Then we introduce a proper $\mathbb{Z}_3$ identification, e.g.
\begin{eqnarray}
\phi_k = \varphi_k + \varphi_{km} + \varphi_{km^2}\,.
\end{eqnarray}
Consequently each field $\phi_k$ corresponds to a class $[k]$ of $\mathbb{Z}_3$ gauging of $\mathbb{Z}_N$ symmetry. In the $\mathbb{Z}_2$ gauging $\mathbb{Z}_N$ symmetry, the field $\phi_k$ and it conjugate $\phi^*_k$ correspond to the same class, but they are complex conjugate each other in $\mathbb{Z}_3$ gauging of $\mathbb{Z}_N$ symmetry. Thus, one can introduce complex representations in a theory with $\mathbb{Z}_3$ gauging of $\mathbb{Z}_N$ symmetry.

\section{Neutrino mass matrix of two-zero texture from fussion rules of $\mathbb{Z}_3$ gauging of $\mathbb{Z}_N$ symmetry \label{sec:neu_mass}}

In this section, we shall investigate the patterns of the neutrino mass matrix which can be obtained from the $\mathbb{Z}_3$ gauging of $\mathbb{Z}_N$ symmetry. In particular, we are concerned with the most predictive textures with two zero elements. We shall consider two scenarios in which neutrino mass is generated by the Weinberg operator or the type-I seesaw mechanism.   

\subsection{Neutrino mass from Weinberg operator}

Let us focus on Majorana neutrinos and its mass is described by the Weinberg operators. The Lagrangian for the mass of charged lepton and neutrinos takes the following form, 
\begin{eqnarray}
- \mathcal{L}^{{\rm lepton}}_{m}= (Y_E)_{ij} \overline{\ell_{Li}}H E_{Rj} + \frac{C_W^{ij}}{\Lambda} \left(\overline{\ell_{Li}}\tilde{H}\right)(\widetilde{H}^T \ell_{Lj}^C)+\text{h.c.}  \,,\label{eq:Lag-charged-lepton-neutrino}
\end{eqnarray}
where $i, j = 1,2,3$ are the flavor indices, $C$ denotes the charge conjugation matrix, and $\Lambda$ is the scale of lepton number violation. Notice that two Higgs doublets $H_u$ and $H_d$ are necessary in the minimal supersymmetric standard model (MSSM), they enter into the neutrino and charged lepton mass terms in Eq.~\eqref{eq:Lag-charged-lepton-neutrino}, respectively. After electroweak symmetry breaking, the charged leptons and neutrinos acquire their masses as follows: 
\begin{eqnarray}
M_E = \dfrac{v}{\sqrt{2}} Y_E \,,~~~ M_\nu = \dfrac{v^2}{\Lambda} C_W \,,
\end{eqnarray}
where $v\simeq 246\,{\rm GeV}$ is the vacuum expectation value of the Higgs fields.
In what follows, we shall perform a comprehensive analysis of the textures of $Y_E$ and $C_W^{ij}$ which can be achieved from the $\mathbb{Z}_3$ gauging of $\mathbb{Z}_N$ symmetry with $N = 13, 19$, all independent transformations of lepton fields and Higgs fields under the non-invertible symmetry are considered. The cases of $N=7, 9, 14$ are discussed in the appendix~\ref{sec:Z3_gauging_of_Z7_Z9}, unfortunately they can not give the desired results.

\subsubsection{\label{subseub:N13-Weinberg}$N = 13$}
We study the case with $N = 13, m=3$, where there are five classes $[0]$, $[1]$, $[2]$, $[4]$ and $[7]$. Another solution with $m=3^2=9$ will give the same results. The corresponding equivalence classes are given by
\begin{eqnarray}
[0] = \{e\} \,,~ 
[1] = \{a,a^3, a^9 \} \,,~
[2] = \{a^2,a^5,a^6 \} \,,~
[4] = \{a^4,a^{10},a^{12} \} \,,~
[7] = \{a^7,a^8,a^{11} \} \,,~~~
\end{eqnarray}
The fusion rules are given by
\begin{eqnarray}
\nonumber && [1] \times [1] = [2] + 2 [4] \,,~~~
[1] \times [2] = [1] + [2] + [7] \,,~~~
[1] \times [4] = 3[0] + [2] + [7] \,,~~~\\
\nonumber && [1] \times [7] = [1] + [4] + [7] \,,~~~ 
 [2] \times [2] = [4] + 2 [7] \,,~~~  
[2] \times [4] = [1] + [2] + [4] \,,\\
\nonumber &&
[2] \times [7] = 3[0] + [1] + [4] \,,~~~
[4] \times [4] = 2 [1] + [3] \,,~~~  
 [4] \times [7] = [2] + [4] + [7] \,,\\
&& [7] \times [7] = 2[2] + [1],~~~[0]\times [n]=[n] \,, \label{eq:FS-N13}
\end{eqnarray}
where $n\in\{0, 1, 2, 4, 7\}$.
The automorphism group of $\mathbb{Z}_{13}$ is ${\rm Aut}(\mathbb{Z}_{13})\cong \mathbb{Z}_{12}$. Under the action of an automorphism, the fusion rules remain invariant. In particular, there exists a $\mathbb{Z}_3$ subgroup whose action that leaves the equivalence classes unchanged. This subgroup constitutes the kernel of the symmetry action.
Thus, the fusion rules have a $\mathbb{Z}_{12}/\mathbb{Z}_3 \cong \mathbb{Z}_4$ symmetry which is generated by $u$ such that 
\begin{eqnarray}
\label{eq:FRSym-N13} u([0]) = [0] \,,~~~ u([1]) = [2] \,,~~~ u([2]) = [4] \,,~~~ u([4]) = [7] \,,~~~ u([7])= [1]\,.
\end{eqnarray}
If two or three generations of left-handed lepton fields are assigned to the same equivalence classes, then at least two rows (columns) have the same texture. As a consequence, the neutrino mass matrix could be of the following form
\begin{eqnarray}
\nonumber &&\begin{pmatrix}
\times & \times & \times \\
\times & \times & \times \\
\times & \times & \times 
\end{pmatrix}\,,~~~ 
\begin{pmatrix}
\times & 0 & 0 \\
0 & \times & \times \\
0 & \times & \times 
\end{pmatrix}\,,~~~ 
\begin{pmatrix}
0 & \times & \times \\
\times & \times & \times \\
\times & \times & \times 
\end{pmatrix}\,,~~~ 
\begin{pmatrix}
0 & 0 & 0 \\
0 & \times & \times \\
0 & \times & \times 
\end{pmatrix}\,,~~~ \\
\label{eq:texture-L1L2same}&& \begin{pmatrix}
\times & \times & \times \\
\times & 0 & 0 \\
\times & 0 & 0 
\end{pmatrix}\,,~~~ 
\begin{pmatrix}
\times & 0 & 0 \\
0 & 0 & 0 \\
0 & 0 & 0 
\end{pmatrix}\,,~~~ 
\begin{pmatrix}
0 & \times & \times \\
\times & 0 & 0 \\
\times & 0 & 0
\end{pmatrix}\,,
\end{eqnarray}
up to simultaneous permutations of rows and columns. Obviously the phenomenologically viable neutrino mass matrix with two texture zero in Eq.~\eqref{eq:ABC-2zero} can not be produced. 

If the three generations of left-handed lepton fields $\ell_L$ are assigned to three different equivalence classes, we have the following ten possible assignments up to permutation:
\begin{eqnarray}
\nonumber \ell_L &\sim &~ ([0], [1], [2]) \,,~~~ ([0], [1], [4]) \,,~~~ ([0], [1], [7]) \,,~~~ ([0], [2], [4]) \,, \\
\nonumber &&~ ([0], [2], [7]) \,,~~~ ([0], [4], [7]) \,,~~~ ([1], [2], [4]) \,,~~~ ([1], [2], [7]) \,, \\
\label{eq:Z13_assignment}&&~ ([1], [4], [7]) \,,~~~ ([2], [4], [7]) \,.
\end{eqnarray}
Due to the $\mathbb{Z}_4$ symmetry of the fusion rules, it is sufficient to assign the Higgs field to the equivalence class labeled by either $[0]$ or $[1]$. 
With the possible assignments of $\ell_L$ in Eq.~\eqref{eq:Z13_assignment} and the fussion rules in Eq.~\eqref{eq:FS-N13}, it is straightforward to determine the texture structure of the neutrino mass matrix. The resulting textures are summarized in table~\ref{tab:Z13_Wein}.
\begin{table}[t!]
\centering
\resizebox{\textwidth}{!}{
\begin{tabular}{|c|c|c||c|c|c|}
\hline\hline
\multirow{2}{*}{$\ell_L$} &  \multicolumn{2}{c||}{$M_\nu$} & \multirow{2}{*}{$\ell_L$}  & \multicolumn{2}{c|}{$M_\nu$}  \\ \cline{2-3} \cline{5-6}
& $H\sim [0]$ & $H\sim [1]$ &  & $H\sim [0]$ & $H\sim [1]$   \\
\hline\hline
$([0], [1], [2])$ & 
$\begin{pmatrix}
\times ~& 0 ~& 0  \\
0 ~& 0 ~& 0  \\
0 ~& 0 ~& 0  
\end{pmatrix}$ & 
$\begin{pmatrix}
0 & \times & 0  \\
\times & 0 & \times  \\
0 & \times & \times  
\end{pmatrix}$ & 
$([0], [4], [7])$ & 
$\begin{pmatrix}
\times & 0 & 0  \\
0 & 0 & 0  \\
0 & 0 & 0  
\end{pmatrix}$ & 
$\begin{pmatrix}
0 & 0 & \times  \\
0 & \times & \times  \\
\times & \times & \times  
\end{pmatrix}$  \\ \hline 
$([0], [1], [4])$ & 
$\begin{pmatrix}
\times & 0 & 0  \\
0 & 0 & \times  \\
0 & \times & 0  
\end{pmatrix}$ & 
$\begin{pmatrix}
0 & \times & 0  \\
\times & 0 & \times  \\
0 & \times & \times  
\end{pmatrix}$ & 
$([1], [2], [4])$ & 
$\begin{pmatrix}
0 & 0 & \times  \\
0 & 0 & 0  \\
\times & 0 & 0  
\end{pmatrix}$ & 
$\begin{pmatrix}
0 & \times & \times  \\
\times & \times & \times  \\
\times & \times & \times  
\end{pmatrix}$  \\ \hline 
$([0], [1], [7])$ & 
$\begin{pmatrix}
\times & 0 & 0  \\
0 & 0 & 0  \\
0 & 0 & 0  
\end{pmatrix}$ & 
$\begin{pmatrix}
0 & \times & \times  \\
\times & 0 & \times  \\
\times & \times & \times  
\end{pmatrix}$ & 
$([1], [2], [7])$ & 
$\begin{pmatrix}
0 & 0 & 0  \\
0 & 0 & \times  \\
0 & \times & 0  
\end{pmatrix}$ & 
$\begin{pmatrix}
0 & \times & \times  \\
\times & \times & \times  \\
\times & \times & \times  
\end{pmatrix}$  \\ \hline 
$([0], [2], [4])$ & 
$\begin{pmatrix}
\times & 0 & 0  \\
0 & 0 & 0  \\
0 & 0 & 0  
\end{pmatrix}$ & 
$\begin{pmatrix}
0 & 0 & 0  \\
0 & \times & \times  \\
0 & \times & \times  
\end{pmatrix}$ & 
$([1], [4], [7])$ & 
$\begin{pmatrix}
0 & \times & 0  \\
\times & 0 & 0  \\
0 & 0 & 0  
\end{pmatrix}$ & 
$\begin{pmatrix}
0 & \times & \times  \\
\times & \times & \times  \\
\times & \times & \times  
\end{pmatrix}$  \\ \hline 
$([0], [2], [7])$ & 
$\begin{pmatrix}
\times & 0 & 0  \\
0 & 0 & \times  \\
0 & \times & 0  
\end{pmatrix}$ & 
$\begin{pmatrix}
0 & 0 & \times  \\
0 & \times & \times  \\
\times & \times & \times  
\end{pmatrix}$ & 
$([2], [4], [7])$ & 
$\begin{pmatrix}
0 & 0 & \times  \\
0 & 0 & 0  \\
\times & 0 & 0  
\end{pmatrix}$ & 
$\begin{pmatrix}
\times & \times & \times  \\
\times & \times & \times  \\
\times & \times & \times  
\end{pmatrix}$  \\ \hline
\end{tabular}}
\caption{\label{tab:Z13_Wein}The textures of the neutrino mass matrix $M_{\nu}$ in $\mathbb{Z}_3$ gauging of $\mathbb{Z}_{13}$ symmetry for different possible assignments of lepton doublets $\ell_L$, where the light neutrino mass is described by the Weinberg operator. 
}
\end{table}
We can see that only the textures $\textbf{A}_{1,2}$, $\textbf{B}_{3,4}$ and $\textbf{C}$ can be obtained. The Higgs field should be in the conjugacy class $H\sim[1]$, the corresponding assignments of the lepton fields are given by
\begin{eqnarray}
&& \textbf{A}_1~:~\begin{pmatrix}
0&0&\times \\
0&\times &\times \\
\times &\times &\times 
\end{pmatrix} ~\text{ for }~ \ell_L \sim ([0], [2], [7])\,,~ ([0], [4], [7])  \\
&& \textbf{A}_2~:~\begin{pmatrix}
0&\times&0 \\
\times &\times &\times \\
0&\times &\times
\end{pmatrix} ~\text{ for }~ \ell_L \sim ([0], [7], [2])\,,~ ([0], [7], [4])  \\
&& \textbf{B}_3~:~\begin{pmatrix}
\times& 0& \times \\
0 & 0&\times \\
\times &\times &\times 
\end{pmatrix}  ~\text{ for }~ \ell_L \sim ([2], [0], [7])\,,~ ([4], [0], [7])  \\
&& \textbf{B}_4~:~\begin{pmatrix}
\times&\times& 0 \\
\times & \times &\times \\
0 &\times &0 
\end{pmatrix}  ~\text{ for }~ \ell_L \sim ([2], [7], [0])\,,~ ([4], [7], [0])  \\
&& \textbf{C}~:~\begin{pmatrix}
\times& \times& \times \\
\times & 0&\times \\
\times &\times &0
\end{pmatrix}  ~\text{ for }~ \ell_L \sim ([7], [0], [1])\,,~ ([7], [1], [0])  \,.
\end{eqnarray}
Now we proceed to consider the charged lepton sector. When we work in the framework of MSSM, there are two Higgs fields $H_u$ and $H_d$ which can correspond to different classes. We can always assign $H_d \sim [0]$,  the right-handed charge leptons $E_{Ri}$ are in the same conjugacy class as $\ell_{Li}$, then the charged lepton Yukawa coupling matrix would be diagonal.

\begin{table}[t!]
\centering
\begin{tabular}{|c|c|c|c|}
\hline\hline
& $E_R \sim ([1],[2],[4])$ & $E_R \sim ([1],[4],[7])$ & $E_R \sim ([2],[4],[7])$ \\ \cline{2-4}  
$\ell_L \sim ([0],[2],[7])$ 
& $\begin{pmatrix}
0 & 0 & \times  \\
\times & \times & \times  \\
0 & \times & \times
\end{pmatrix}$
& $\begin{pmatrix}
0 & \times & 0  \\
\times & \times & 0  \\
0 & \times & \times
\end{pmatrix}$
& $\begin{pmatrix}
0 & \times & 0  \\
\times & \times & 0  \\
\times & \times & \times
\end{pmatrix}$  \\ \hline
 & $E_R \sim ([1],[2],[4])$ & $E_R \sim ([1],[4],[7])$ & $E_R \sim ([2],[4],[7])$ \\ \cline{2-4}  
$\ell_L \sim ([0],[4],[7])$ 
& $\begin{pmatrix}
0 & 0 & \times  \\
\times & 0 & 0  \\
0 & \times & \times
\end{pmatrix}$
& $\begin{pmatrix}
0 & \times & 0  \\
\times & 0 & \times  \\
0 & \times & \times
\end{pmatrix} $
& $\begin{pmatrix}
0 & \times & 0  \\
0 & 0 & \times  \\
\times & \times & \times
\end{pmatrix}$   \\ \hline
 & $E_R \sim ([0],[2],[4])$ & $([0],[4],[7])$ & $([2],[4],[7])$ \\ \cline{2-4}  
$\ell_L \sim ([0],[1],[7])$
& $\begin{pmatrix}
0 & 0 & \times  \\
\times & \times & 0  \\
0 & \times & \times
\end{pmatrix}$
& $\begin{pmatrix}
0 & \times & 0  \\
\times & 0 & \times  \\
0 & \times & \times
\end{pmatrix}$ 
& $\begin{pmatrix}
0 & \times & 0  \\
\times & 0 & \times  \\
\times & \times & \times
\end{pmatrix}$  \\ \hline\hline
\end{tabular}
\caption{\label{tab:Z13_ME}The texture of the full ranked charged lepton Yukawa matrices $M_E$ in $\mathbb{Z}_3$ gauging of $\mathbb{Z}_{13}$ symmetry for different possible assignments of right-handed charged leptons $E_R$, in which cases the neutrino mass matrix has two-zero texture and the Higgs is assigned to $[1]$.  }
\end{table}
If we work in the SM which has only one Higgs doublet, the Higgs field must be assigned as $H \sim [1]$ to realize two texture zeros in the neutrino mass matrix, as explained above. Here, for the assignments of left-handed charged leptons that lead to two texture zeros, we list the corresponding assignments of right-handed charged leptons that render the charged lepton mass matrix full rank. Up to permutations, the assignments of the lepton fields and the corresponding charged lepton mass matrix textures are given in table~\ref{tab:Z13_ME}. We see that the charged lepton Yukawa matrices is not be diagonal in this scenario.

\subsubsection{$N=19$}\label{sec:weinbop-N19}

We study the case with $N = 19, m=7$, where there are seven equivalence classes $[0]$, $[1]$, $[2]$, $[4]$, $[5]$, $[8]$ and $[10]$ which are given by~\footnote{Another solution with $m=11$ will give the same set of equivalence classes.} 
\begin{eqnarray}
\nonumber && [0] = \{ e \} \,,~~~ 
[1] = \{ a, a^7, a^{11} \} \,,~~~
[2] = \{ a^2, a^3, a^{14} \} \,,~~~
[4] = \{ a^4, a^6, a^9 \} \,, \\
&& [5] = \{ a^5, a^{16}, a^{17} \} \,,~~~
[8] = \{ a^8, a^{12}, a^{18} \} \,,~~~
[10] = \{ a^{10}, a^{13}, a^{15} \} \,.
\end{eqnarray}
Their fusion rules are listed in table~\ref{tab:Z19_fusion}.
\begin{table}[b!]
\centering
\resizebox{1.01\textwidth}{!}{
\begin{tabular}{|c||c|c|c|c|c|c|c|}
\hline\hline
\diagbox{$[k]$}{$[k]\times [l]$}{$[l]$} & $[0]$ & $[1]$ & $[2]$ & $[4]$ & $[5]$ & $[8]$ & $[10]$  \\ \hline\hline
$[0]$ & $[0]$ & $[1]$ & $[2]$ & $[4]$ & $[5]$ & $[8]$ & $[10]$  \\ \hline
$[1]$ & $[1]$ & $[2] + 2[8]$ & $[2] + [4] + [10]$ & $[1] + [5] + [10]$ & $[4] + [5] + [8]$ & $3[0] + [4] + [10]$ & $[1] + [2] + [5]$ \\ \hline
$[2]$ & $[2]$ & $[2] + [4] + [10]$ & $[4] + 2[5]$ & $[1] + [4] + [8]$ & $3[0] + [1] + [8]$ & $[1] + [2] + [10]$ & $[5] + [8] + [10]$  \\ \hline
$[4]$ & $[4]$ & $[1] + [5] + [10]$ & $[1] + [4] + [8]$ & $[8] + 2[10]$ & $[1] + [2] + [4]$ & $[2] + [5] + [8]$ & $3[0] + [2] + [5]$ \\ \hline
$[5]$ & $[5]$ & $[4] + [5] + [8]$ & $3[0] + [1] + [8]$ & $[1] + [2] + [4]$ & $2[2] + [10]$ & $[4] + [5] + [10]$ & $[1] + [8] +[10]$ \\ \hline
$[8]$ & $[8]$ & $3[0] + [4] + [10]$ & $[1] + [2] + [10]$ & $[2] + [5] + [8]$ & $[4] + [5] + [10]$ & $2[1] + [5]$ & $[2] + [4] + [8]$  \\ \hline
$[10]$ & $[10]$ & $[1] + [2] + [5]$ & $[5] + [8] + [10]$ & $3[0] + [2] + [5]$ & $[1] + [8] +[10]$ & $[2] + [4] + [8]$ & $[1] + 2[4]$  \\ \hline\hline
\end{tabular}}
\caption{\label{tab:Z19_fusion} The fusion rules for the $\mathbb{Z}_3$ gauging of $\mathbb{Z}_{19}$ symmetry. }
\end{table}
Similar to the case with $N=13$ as discussed in the previous section, the automorphism group of $\mathbb{Z}_{19}$ is ${\rm Aut}(\mathbb{Z}_{19}) \cong \mathbb{Z}_{18}$.
The fusion rules have a $\mathbb{Z}_{18}/\mathbb{Z}_3 \cong \mathbb{Z}_6$ symmetry generated by 
\begin{eqnarray}
\nonumber && u([0]) = [0] \,,~~~
u([1]) = [2] \,,~~~
u([2]) = [4] \,,~~~
u([4]) = [8] \,,\\
&& u([5]) = [10] \,,~~~
u([8]) = [5]  \,,~~~
u[(10)] = [1] \,.~~~
\end{eqnarray}
Therefore it is sufficient to only consider the assignments $H\sim [0]$ and $H\sim [1]$.

Let us first consider the assignment $H\sim [0]$. If the three generations of leptons are assigned to different classes, they can be assigned as $([0], [i], [j])$ or $([l], [i], [j])$ with $l\neq i\neq j$. In our cases, we always have $[0] \nsubseteq [i]\times [i]$ and $[0] \times [i] = [i]$. Since the three generations of left-handed leptons are assigned to different classes, at most one contraction out of $[l]\times [i]$, $[l]\times [j]$ and $[i]\times [j]$ can contain identity class $[0]$. Without loss of generality, we can  take $[0] \nsubseteq [l] \times [i]$ and $[0] \nsubseteq [l] \times [j]$. Hence only the following four patterns of the neutrino mass matrix can be achieve,
\begin{eqnarray}
\label{eq:H0-nuMM1}\ell_L \sim ([0], [i], [j]), [0] \subseteq [i] \times [j] ~&:&~~
\begin{pmatrix}
\times ~& 0 ~& 0 \\
0 ~& 0 ~& \times \\
0 ~& \times ~& 0 
\end{pmatrix}\,,\\
\label{eq:H0-nuMM2}\ell_L \sim ([0], [i], [j]), [0] \nsubseteq [i] \times [j] ~&:&~~
\begin{pmatrix}
\times ~& 0 ~& 0 \\
0 ~& 0 ~& 0 \\
0 ~& 0 ~& 0 
\end{pmatrix}\,,  \\
\label{eq:H0-nuMM3} \ell_L \sim ([l], [i], [j]), [0] \subseteq [i] \times [j] ~&:&~~
\begin{pmatrix}
0 ~& 0 ~& 0 \\
0 ~& 0 ~& \times \\
0 ~& \times ~& 0 
\end{pmatrix}\,, \\
\label{eq:H0-nuMM4}\ell_L \sim ([l], [i], [j]), [0] \nsubseteq [i] \times [j] ~&:&~~
\begin{pmatrix}
0 ~& 0 ~& 0 \\
0 ~& 0 ~& 0 \\
0 ~& 0 ~& 0 
\end{pmatrix} \,,
\end{eqnarray}
up to row and column permutations. We see that 
the neutrino mass matrices contain at least four texture zero which is not phenomenologically viable.
Consequently, we do not consider the assignment $H\sim [0]$, and we focus exclusively on the possible forms of neutrino mass matrix corresponding to the assignment $H\sim [1]$, as summarized in table~\ref{tab:Z19_Wein}.

\begin{table}[h!]
\centering
\resizebox{\textwidth}{!}{
\begin{tabular}{|c|c||c|c||c|c||c|c|}
\hline\hline
$\ell_L$ & $M_\nu$ & $\ell_L$ & $M_\nu$ & $\ell_L$ & $M_\nu$ & $\ell_L$ & $M_\nu$   \\ \hline\hline
$([0], [1], [2])$ & 
$\begin{pmatrix}
0 & \times & 0  \\
\times & 0 & 0  \\
0 & 0 & \times  
\end{pmatrix}$ & 
$([0], [4], [5])$ & 
$\begin{pmatrix}
0 & 0 & \times  \\
0 & 0 & \times  \\
\times & \times & 0  
\end{pmatrix}$ & 
$([1], [2], [10])$ & 
$\begin{pmatrix}
0 & 0 & \times  \\
0 & \times & \times  \\
\times & \times & \times  
\end{pmatrix}$ & 
$([2], [4], [10])$ & 
$\begin{pmatrix}
\times & \times & \times  \\
\times & 0 & \times  \\
\times & \times & \times  
\end{pmatrix}$  \\ \hline 
$([0], [1], [4])$ & 
$\begin{pmatrix}
0 & \times & 0  \\
\times & 0 & \times  \\
0 & \times & 0  
\end{pmatrix}$ & 
$([0], [4], [8])$ & 
$\begin{pmatrix}
0 & 0 & 0  \\
0 & 0 & \times  \\
0 & \times & \times  
\end{pmatrix}$ & 
$([1], [4], [5])$ & 
$\begin{pmatrix}
0 & \times & \times  \\
\times & 0 & \times  \\
\times & \times & 0  
\end{pmatrix}$ & 
$([2], [5], [8])$ & 
$\begin{pmatrix}
\times & \times & \times  \\
\times & 0 & \times  \\
\times & \times & \times  
\end{pmatrix}$  \\ \hline 
$([0], [1], [5])$ & 
$\begin{pmatrix}
0 & \times & \times  \\
\times & 0 & \times  \\
\times & \times & 0  
\end{pmatrix}$ & 
$([0], [4], [10])$ & 
$\begin{pmatrix}
0 & 0 & 0  \\
0 & 0 & \times  \\
0 & \times & \times  
\end{pmatrix}$ & 
$([1], [4], [8])$ & 
$\begin{pmatrix}
0 & \times & 0  \\
\times & 0 & \times  \\
0 & \times & \times  
\end{pmatrix}$ & 
$([2], [5], [10])$ & 
$\begin{pmatrix}
\times & \times & \times  \\
\times & 0 & \times  \\
\times & \times & \times  
\end{pmatrix}$  \\ \hline 
$([0], [1], [8])$ & 
$\begin{pmatrix}
0 & \times & 0  \\
\times & 0 & 0  \\
0 & 0 & \times  
\end{pmatrix}$ & 
$([0], [5], [8])$ & 
$\begin{pmatrix}
0 & \times & 0  \\
\times & 0 & \times  \\
0 & \times & \times  
\end{pmatrix}$ & 
$([1], [4], [10])$ & 
$\begin{pmatrix}
0 & \times & \times  \\
\times & 0 & \times  \\
\times & \times & \times  
\end{pmatrix}$ & 
$([2], [8], [10])$ & 
$\begin{pmatrix}
\times & \times & \times  \\
\times & \times & 0  \\
\times & 0 & \times  
\end{pmatrix}$  \\ \hline 
$([0], [1], [10])$ & 
$\begin{pmatrix}
0 & \times & 0  \\
\times & 0 & \times  \\
0 & \times & \times  
\end{pmatrix}$ & 
$([0], [5], [10])$ & 
$\begin{pmatrix}
0 & \times & 0  \\
\times & 0 & \times  \\
0 & \times & \times  
\end{pmatrix}$ & 
$([1], [5], [8])$ & 
$\begin{pmatrix}
0 & \times & 0  \\
\times & 0 & \times  \\
0 & \times & \times  
\end{pmatrix}$ & 
$([4], [5], [8])$ & 
$\begin{pmatrix}
0 & \times & \times  \\
\times & 0 & \times  \\
\times & \times & \times  
\end{pmatrix}$  \\ \hline 
$([0], [2], [4])$ & 
$\begin{pmatrix}
0 & 0 & 0  \\
0 & \times & \times  \\
0 & \times & 0  
\end{pmatrix}$ & 
$([0], [8], [10])$ & 
$\begin{pmatrix}
0 & 0 & 0  \\
0 & \times & 0  \\
0 & 0 & \times  
\end{pmatrix}$ & 
$([1], [5], [10])$ & 
$\begin{pmatrix}
0 & \times & \times  \\
\times & 0 & \times  \\
\times & \times & \times  
\end{pmatrix}$ & 
$([4], [5], [10])$ & 
$\begin{pmatrix}
0 & \times & \times  \\
\times & 0 & \times  \\
\times & \times & \times  
\end{pmatrix}$  \\ \hline 
$([0], [2], [5])$ & 
$\begin{pmatrix}
0 & 0 & \times  \\
0 & \times & \times  \\
\times & \times & 0  
\end{pmatrix}$ & 
$([1], [2], [4])$ & 
$\begin{pmatrix}
0 & 0 & \times  \\
0 & \times & \times  \\
\times & \times & 0  
\end{pmatrix}$ & 
$([1], [8], [10])$ & 
$\begin{pmatrix}
0 & 0 & \times  \\
0 & \times & 0  \\
\times & 0 & \times  
\end{pmatrix}$ & 
$([4], [8], [10])$ & 
$\begin{pmatrix}
0 & \times & \times  \\
\times & \times & 0  \\
\times & 0 & \times  
\end{pmatrix}$  \\ \hline 
$([0], [2], [8])$ & 
$\begin{pmatrix}
0 & 0 & 0  \\
0 & \times & \times  \\
0 & \times & \times  
\end{pmatrix}$ & 
$([1], [2], [5])$ & 
$\begin{pmatrix}
0 & 0 & \times  \\
0 & \times & \times  \\
\times & \times & 0  
\end{pmatrix}$ & 
$([2], [4], [5])$ & 
$\begin{pmatrix}
\times & \times & \times  \\
\times & 0 & \times  \\
\times & \times & 0  
\end{pmatrix}$ & 
$([5], [8], [10])$ & 
$\begin{pmatrix}
0 & \times & \times  \\
\times & \times & 0  \\
\times & 0 & \times  
\end{pmatrix}$  \\ \hline 
$([0], [2], [10])$ & 
$\begin{pmatrix}
0 & 0 & 0  \\
0 & \times & \times  \\
0 & \times & \times  
\end{pmatrix}$ & 
$([1], [2], [8])$ & 
$\begin{pmatrix}
0 & 0 & 0  \\
0 & \times & \times  \\
0 & \times & \times  
\end{pmatrix}$ & 
$([2], [4], [8])$ & 
$\begin{pmatrix}
\times & \times & \times  \\
\times & 0 & \times  \\
\times & \times & \times  
\end{pmatrix}$ & &  \\ \hline \hline
\end{tabular}}
\caption{\label{tab:Z19_Wein}
The texture of neutrino mass matrix $M_{\nu}$ in $\mathbb{Z}_3$ gauging of $\mathbb{Z}_{19}$ symmetry for different possible assignments of lepton doublets $\ell_L$, where the light neutrino mass is described by the Weinberg operator and the Higgs field transforms as $H\sim[1]$. }
\end{table}

We see that all the seven textures $\textbf{A}_{1,2}$, $\textbf{B}_{1,2,3,4}$ and $\textbf{C}$ in~\cite{Fritzsch:2011qv} can be obtained. 
These textures together with diagonal charged lepton mass matrix can be generated from the non-invertible symmetry for the following assignments with only one Higgs doublet $H\sim[1]$.
\begin{eqnarray}
&& \textbf{A}_1~:~\begin{pmatrix}
0&0&\times \\
0&\times &\times \\
\times &\times &\times 
\end{pmatrix} ~\text{ for }~ (\ell_L; E_R) \sim ([1], [2], [10]; [0], [1], [8]) \,,  \\
&& \textbf{A}_2~:~\begin{pmatrix}
0&\times&0 \\
\times &\times &\times \\
0&\times &\times
\end{pmatrix} ~\text{ for }~ (\ell_L; E_R) \sim ([1], [10], [2]; [0], [8], [1])  \,,  \\
&& \textbf{B}_1~:~\begin{pmatrix}
\times&\times& 0 \\
\times & 0&\times \\
0 &\times &\times 
\end{pmatrix} ~\text{ for }~ (\ell_L; E_R) \sim ([8], [5], [10]; [1], [10], [2])\,,~ ([10], [5], [8]; [2], [10], [1])\,,  \\
&& \textbf{B}_2~:~\begin{pmatrix}
\times&0& \times \\
0 &\times &\times \\
\times &\times &0 
\end{pmatrix} ~\text{ for }~ (\ell_L; E_R) \sim ([8], [10], [5]; [1], [2], [10])\,,~ ([10], [8], [5]; [2], [1], [10]) \,, \\
&& \textbf{B}_3~:~\begin{pmatrix}
\times& 0& \times \\
0 & 0&\times \\
\times &\times &\times 
\end{pmatrix} ~\text{ for }~ (\ell_L; E_R) \sim ([2], [1], [10]; [1], [0], [8]) \,, \\
&& \textbf{B}_4~:~\begin{pmatrix}
\times&\times& 0 \\
\times & \times &\times \\
0 &\times &0 
\end{pmatrix} ~\text{ for }~ (\ell_L; E_R) \sim ([2], [10], [1]; [1], [8], [0]) \,, \\
\label{eq:Yukawa_N19_C}&& \textbf{C}~:~\begin{pmatrix}
\times& \times& \times \\
\times & 0&\times \\
\times &\times &0
\end{pmatrix}\begin{array}{ll}
~\text{ for }~ (\ell_L; E_R) \sim  & ([10], [1], [5]; [8], [0], [5]) \,,~ ([10], [5], [1]; [8], [5], [0]) \\ 
& ([2], [4], [5]; [1], [8], [4]) \,,~ ([2], [5], [4]; [1], [4], [8])  \\ 
& ([8], [4], [5]; [1], [2], [4]) \,,~ ([8], [5], [4]; [1], [4], [2])  \\ 
& ([8], [4], [5]; [1], [2], [10]) \,,~ ([8], [5], [4]; [1], [10], [2])  
\end{array}\,.
\end{eqnarray}
The charged lepton mass matrix is diagonal for the above assignments.  
If the model contains only one Higgs field, the other cases that the neutrino masses have two texture zero cannot have diagonal charged lepton Yukawa couplings. 
With the introduction of two Higgs fields, these additional assignments for the left-handed leptons can simultaneously yield a neutrino mass matrix with two texture zeros and a diagonal charged lepton mass matrix. As discussed in the cases with $N=13$, assigning the Higgs field $H_d$ to the equivalence class $[0]$ is one viable option. Here we present several other nontrivial assignments as examples, which also lead to a diagonal charged lepton mass matrix. 
\begin{eqnarray}
\nonumber && \textbf{B}_1~:~\begin{pmatrix}
\times&\times& 0 \\
\times & 0&\times \\
0 &\times &\times 
\end{pmatrix} ~\text{ for }~ (\ell_L; E_R; H_d)  \sim ([8], [4], [10]; [5], [2], [8]; [2])\,,~ ([10], [4], [8]; [1], [0], [8]; [4])\,,  \\
\nonumber && \textbf{B}_2~:~\begin{pmatrix}
\times&0& \times \\
0 &\times &\times \\
\times &\times &0 
\end{pmatrix} ~\text{ for }~ (\ell_L; E_R; H_d) \sim ([8], [10], [4]; [8], [1], [5]; [4])\,,~ ([10], [8], [4]; [5], [2], [4]; [5])\,,  \\
\nonumber && \textbf{C}~:~\begin{pmatrix}
\times& \times& \times \\
\times & 0&\times \\
\times &\times &0
\end{pmatrix}\begin{array}{ll} ~\text{ for }~ (\ell_L; E_R; H_d) \sim & ([10], [1], [4]; [10], [5], [2; [2]]) \,,~ ([10], [4], [1]; [5], [1], [2]; [5])\,,   \\
&([10], [4], [5]; [2], [10], [4]; [8]) \,,~ ([10], [5], [4]; [0], [1], [8]; [10])
\end{array}  \,,
\end{eqnarray}
for which the charged lepton mass matrix is diagonal.

\subsection{Neutrino masses from seesaw mechanism \label{sec:seesaw}}

We now consider the scenario that light neutrino mass arises from the type-I seesaw mechanism.
The SM gauge invariant Lagrangian for the neutrino masses can be written as
\begin{eqnarray}
\label{eq:seesaw-Lagr}- \mathcal{L}_{\nu} = (Y_{\nu})_{ij} \overline{\ell_{Li}} \widetilde{H} \nu_{Rj} + \dfrac{1}{2} (M_N)_{\alpha\beta} \overline{\nu^C_{R\alpha}} \nu_{R\beta} +\text{h.c.}\,, 
\end{eqnarray}
where $i, j,\alpha, \beta=1, 2, 3$ are generation indices, $\nu_{Ri}$ with $i=1, 2, 3$ denote the right-handed neutrino fields, and $\widetilde{H} = i\sigma_2 H^*$.
The Dirac neutrino Yukawa couplings and right-handed neutrino mass matrices are described by $Y_{\nu}$ and $M_N$. Integrating out the right-handed neutrinos, the effective Majorana neutrino mass matrix after electroweak symmetry breaking, is given by the seesaw formula
\begin{eqnarray}
M_\nu = - \dfrac{1}{2} v^2 Y_{\nu} M_N^{-1} Y_{\nu}^T  \,,
\end{eqnarray}
Before analyzing concrete models based on specific $\mathbb{Z}_N$ fusion rules gauging by $\mathbb{Z}_3$, we first present a general discussion. We will show that the texture $\textbf{A}_{1,2}$ and $\textbf{B}_{1,2,3,4}$ cannot be realized within this framework.

It was shown in~\cite{Kaidi:2024wio,Heckman:2024obe,Funakoshi:2024uvy} that the selection rules are preserved at tree level. 
Specifically, if the neutrino mass generated by type I seesaw has non-zero entry with $(M_\nu)_{ij}\neq 0$, i.e. 
\begin{eqnarray}
[0] \subset [k_{\ell_{Li}}] \times [k_H] \times [-k_{\nu_{R\alpha}}] \,,~~~
[0] \subset [k_{\ell_{Lj}}] \times [k_H] \times [-k_{\nu_{R\beta}}] \,,~~~
[0] \subset [k_{\nu_{R\alpha}}] \times [k_{\nu_{R\beta}}]\,.
\end{eqnarray}
There exist $\widetilde{k}_{\ell_{Li}}\in [k_{\ell_{Li}}]$, $\widetilde{k}_H,\widetilde{k}'_H \in [k_H]$ and $\widetilde{k}_{\nu_{R\alpha}}, \widetilde{k}'_{\nu_{R\alpha}} \in [k_{\nu_{R\alpha}}]$ such that
\begin{eqnarray}
0 = \widetilde{k}_{\ell_{Li}} + \widetilde{k}_H - \widetilde{k}_{\nu_{R\alpha}} \,,~~~
0 = \widetilde{k}_{\ell_{Lj}} + \widetilde{k}'_H - \widetilde{k}_{\nu_{R\beta}} \,,~~~
\widetilde{k}'_{\nu_{R\alpha}} = - \widetilde{k}'_{\nu_{R\beta}} ~~({\rm mod}~N)  \,. \label{eq:constr-class-seesaw}
\end{eqnarray}
Here $[k_{\psi}]$ denotes the equivalence class assigned to the field $\psi$ and $\widetilde{k}_{\psi}$ stands for any element of the class $[k_{\psi}]$. The last equality of Eq.~\eqref{eq:constr-class-seesaw} implies $\widetilde{k}_{\nu_{R\alpha}} = - m^x \widetilde{k}_{\nu_{R\beta}}$ for some integer $x$. Then 
\begin{eqnarray}
0 = \widetilde{k}_{\ell_{Li}} + \widetilde{k}_H - \widetilde{k}_{\nu_{R\alpha}} + m^x (\widetilde{k}_{\ell_{Lj}} + \widetilde{k}'_H - \widetilde{k}_{\nu_{R\beta}}) 
= \widetilde{k}_{\ell_{Li}} + \widetilde{k}_H + m^x \widetilde{k}_{\ell_{Lj}} + m^x \widetilde{k}'_H  ~~({\rm mod}~N) \,,
\end{eqnarray}
which implies
\begin{eqnarray}
[0] \subset [k_{\ell_{Li}}] \times [k_{\ell_{Lj}}] \times [k_H] \times [k_H] \,.
\end{eqnarray}
Therefore, if the assignments of the left-handed leptons and the Higgs field under the $\mathbb{Z}_3$ gauging of $\mathbb{Z}_N$ symmetry are the same in the seesaw mechanism and the Weinberg operator description of neutrino mass, then any zero entry arising in the neutrino mass matrix for the Weinberg operator necessarily implies a zero in the corresponding matrix element obtained via the seesaw mechanism.
Hence we shall not consider the assignment $H\sim [0]$, since the light neutrino mass matrix would contain at least four texture zeros and it is not viable, as shown  in Eqs.~(\ref{eq:H0-nuMM1}, \ref{eq:H0-nuMM2}, \ref{eq:H0-nuMM3}, \ref{eq:H0-nuMM4}).

In the following, we focus on the case that $N$ is a prime integer. 
As discussed before, the right-handed neutrino mass of rank three can only be of the following form,
\begin{eqnarray}
\nonumber&&\nu_R \sim ([0],[k],[-k]):~
M_N = \begin{pmatrix}
\times & 0 & 0  \\
0 & 0 & \times  \\
0 & \times & 0
\end{pmatrix}\,,~\\
&&\nu_R \sim ([0],[0],[0]):~
M_N = \begin{pmatrix}
\times & \times & \times  \\
\times & \times & \times  \\
\times & \times & \times
\end{pmatrix}\,,~~
\end{eqnarray}
with $k\neq0$. The absence of zero entries in $M_N$ generally implies that $M_{\nu}$ will also have no vanishing elements without cancellation or fine-tuning, unless one row of 
$Y_{\nu}$ is zero, causing the light neutrino mass matrix to become block-diagonal. Thus, we focus on the case that the right-handed neutrinos are assigned to $([0],[k],[-k])$ in the following. 

In this case, the texture $\textbf{B}_{1,2}$ cannot be produced by type-I seesaw mechanism with three right-handed neutrinos, unless there are some nontrivial correlations between different entries of the neutrino Yukawa couplings $Y_{\nu}$. The texture $\textbf{A}_{1,2}$ and $\textbf{B}_{3,4}$ can be realized if the Yukawa couplings take some specific textures. We take the texture $\textbf{A}_{1}$ as an example, and it can be achieved if the neutrino Yukawa coupling $Y_{\nu}$ takes the following patterns:
\begin{eqnarray}\label{eq:seesaw_A1_Yukawa}
Y_\nu =\begin{pmatrix}
0 & 0 & \times  \\
\times & 0 & 0  \\
\times & \times & 0
\end{pmatrix} \,,~~~
\begin{pmatrix}
0 & 0 & \times  \\
\times & 0 & 0  \\
\times & \times & \times 
\end{pmatrix} \,,~~~
\begin{pmatrix}
0 & 0 & \times  \\
\times & 0 & \times  \\
\times & \times & 0
\end{pmatrix} \,,~~~
\begin{pmatrix}
0 & 0 & \times  \\
\times & 0 & \times  \\
\times & \times & \times
\end{pmatrix} \,,~~\text{or}~~
\begin{pmatrix}
0 & 0 & \times  \\
\times & 0 & \times  \\
0 & \times & \times
\end{pmatrix} \,,
\end{eqnarray}
including the permutation of the second and third columns. In all of these textures, we find $(Y_\nu)_{11} = (Y_\nu)_{12} = (Y_\nu)_{22} = 0$ and $(Y_\nu)_{13}, (Y_\nu)_{21}, (Y_\nu)_{32} \neq 0 $. The Higgs field is assigned to the class [1], i.e. $H\sim[1]$. Supposing the three generations of left-handed leptons are assigned to three different equivalence classes $[l_1]$, $[l_2]$ and $[l_3]$ with $[l_1]\neq[l_2]\neq[l_3]$. The condition $(Y_\nu)_{21} \neq 0$ requires
\begin{eqnarray}
[0] \subset [-l_2] \times [-1]\times [0] \,, 
\end{eqnarray}
which leads to $[l_2] = [-1]$. Therefore we have $[l_3] \neq [-1]$, which implies $(Y_\nu)_{31} = 0$. Thus only the last texture in Eq.~\eqref{eq:seesaw_A1_Yukawa} survives, and it has $(Y_{\nu})_{22}= 0$ and $(Y_{\nu})_{23} \neq 0$ which require
\begin{eqnarray}
[0] \not\subset [-l_2] \times [k] \times  [-1]\,,~~~[0] \subset [-l_2] \times [-k] \times [-1]\,.
\end{eqnarray}
Nevertheless these two condition cannot be fulfilled simultaneously. Therefore the Yukawa coupling textures in Eq.~\eqref{eq:seesaw_A1_Yukawa} as well as the texture $\textbf{A}_1$ cannot be generated from $\mathbb{Z}_3$ gauging of $\mathbb{Z}_N$ fusion rules. The textures $\textbf{A}_2$ and $\textbf{B}_{3,4}$ are related to $\textbf{A}_1$ by exchanging the left-handed leptons~\footnote{By exchanging the assignments of $\ell_{L2}$ and $\ell_{L3}$, the texture $\textbf{A}_2$ can be obtained from texture $\textbf{A}_1$. Similarly, exchanging the assignments of $\ell_{L1}$ and $\ell_{L2}$ yields texture $\textbf{B}_3$, while performing the cyclic permutation $\ell_{L1} \rightarrow \ell_{L3} \rightarrow \ell_{L2}$ leads to texture $\textbf{B}_4$.}, consequently they also cannot be generated from the $\mathbb{Z}_3$ gauging of $\mathbb{Z}_N$ symmetry  in the framework of type-I seesaw mechanism.

Now we consider the Yukawa coupling textures which can generate the texture $\textbf{C}$ of light neutrino mass matrix. The Yukawa couplings can only be of the following form
\begin{eqnarray}\label{eq:seesaw_C_Yukawa}
Y_\nu = 
\begin{pmatrix}
0 & \times & \times \\
0 & 0 & \times  \\
0 & \times & 0 
\end{pmatrix} ~~~\text{ or }~~~
\begin{pmatrix}
\times & \times & \times \\
0 & 0 & \times  \\
0 & \times & 0 
\end{pmatrix} \,, \label{eq:Ynu-textureC}
\end{eqnarray}
up to the permutation of the second the third columns. For the first texture of $Y_{\nu}$ in Eq.~\eqref{eq:Ynu-textureC}, one right-handed neutrino is decoupled so that the lightest neutrino is massless. 
Although the texture $\textbf{C}$ can be compatible with experimental data in general, it is not viable when one of the light neutrino masses vanishes~\cite{Fritzsch:2011qv,Barreiros:2018ndn,treesukrat:2025dhd}. Consequently we only consider the second Yukawa texture in Eq.~\eqref{eq:seesaw_C_Yukawa} whose first row is non-vanishing. Hence the following conditions must be satisfied,  
\begin{eqnarray}
[0] \subset  [-l_1] \times [0] \times [-1] \,,~~~ [0] \subset [-l_1] \times[k] \times [-1]\,,~~~ [0] \subset [-l_1] \times[-k] \times [-1] \,,
\end{eqnarray}
from which we obtain 
\begin{eqnarray}
[l_1] = [-1] ~~~\text{ and } ~~~ [k] = [m-1] \text{ or } [1-m]  \,.
\end{eqnarray}
Without loss of generality, one can choose $[k] = [m-1]$. In the second row of $Y_{\nu}$, only the (23) entry is nonzero and this requires 
\begin{eqnarray}
[0] \not\subset [-l_{2}] \times[0] \times  [-1] \,,~~~
[0] \not\subset [-l_2] \times[k] \times  [-1] \,,~~~ [0] \subset [-l_2] \times [-k] \times  [-1]\,,
\end{eqnarray}
which gives rise to 
\begin{eqnarray}
[l_2] = [1-2m] ~~~\text{ with }~~~ [1-2m] \neq [-1] \text{ and } [2]  \text{ and } [m-2]  \,.
\end{eqnarray}
Similarly considering the third row of $Y_{\nu}$, we can determine the assignment $[l_3]$ as follows,
\begin{eqnarray}
[l_3] = [m-2] ~~~\text{ with }~~~ [m-2] \neq [-1] \text{ and } [2]  \text{ and } [1-2m]  \,.
\end{eqnarray}
Obviously the conditions $[1-2m] \neq [-1]$, $[m-2] \neq [-1]$ and $[1-2m] \neq [m-2]$ is fulfilled for $m\neq1$. However, the condition  $[1-2m] \neq [2]$ or $[m-2]\neq[2]$ can be satisfied when $N=7$~\footnote{In the case of $N=7$, we have $[1-2m]=[2]$ and $[m-2]=[0]$ for $m=2$, whereas $[m-2]=[2]$ and $[1-2m]=[0]$ for $m=4$.}. Hence no neutrino mass matrix with two-zero texture can be obtained from the $\mathbb{Z}_3$ gauging of $\mathbb{Z}_{7}$ symmetry when neutrino mass is generated by type I seesaw mechanism with three right-handed neutrinos. If the prime number $N > 7$, only the texture $\textbf{C}$ can be obtained in seesaw mechanism. Accordingly the assignments of the leptons and Higgs fields are
\begin{eqnarray}
\nonumber (\ell_L; \nu_R; H) &\sim & ([-1], [1-2m], [m-2]; [0], [m-1], [1-m]; [1])\,, \\
&& ([-1], [m-2], [1-2m]; [0], [m-1], [1-m]; [1])  \,. \label{eq:C-seesaw-asign}
\end{eqnarray}
In the following, we give some concrete examples leading to the texture $\textbf{C}$ for $N=13$ and $N=19$.

\subsubsection{$N = 13$}

For $N=13$, $m=3$, we have $[-1] = [4]$, $[m-1] = [2]$, $[m-2] = [1]$ and $[1-2m] = [7]$.
Only texture $\textbf{C}$ can be obtained, and the lepton and Higgs fields are assigned to 
\begin{eqnarray}
\textbf{C}~:~~ (\ell_L; \nu_R; H) ~&\sim &~ ([4], [1], [7]; [0], [2], [7]; [1]) \,,~~ ([4], [7], [1]; [0], [2], [7] ; [1])\,.  
\end{eqnarray}
It should be noticed that, if neutrino mass is described by Weinberg operator, the light neutrino mass matrix has only one texture zero in the (22) or (33) entry for the above assignment. Similar to section~\ref{subseub:N13-Weinberg}, the diagonal charged lepton Yukawa couplings can be achieved if the charged leptons and neutrinos couple to different Higgs fields such as in MSSM.

For completeness, all textures of $M_{\nu}$ that can be obtained from  $\mathbb{Z}_3$ gauging of $\mathbb{Z}_{13}$ symmetry in type I seesaw mechanism are listed in table~\ref{tab:Z13_Seesaw}. 
Among the remaining cases, certain lepton field assignments can also reproduce the texture $\textbf{C}$. However, as discussed previously, these assignments lead to one column of the Dirac neutrino mass matrix being zero, thereby forcing the lightest neutrino mass to be zero. Note that the texture $\textbf{C}$ can not accommodate a vanishing lightest neutrino for the present neutrino oscillation data~\cite{Fritzsch:2011qv,Barreiros:2018ndn,treesukrat:2025dhd}. The corresponding assignments for lepton and Higgs fields are given as follow:
\begin{eqnarray}
\nonumber \textbf{C}~:~~ (\ell_L; \nu_R; H) ~&\sim &~  ([7], [1], [2]; [0], [1], [4]; [1]) \,,~ ([7], [2], [1]; [0], [1], [4]; [1]) \,,  \\
\nonumber &&~ ([2], [1], [7]; [0], [2], [7]; [1]) \,,~ ([2], [7], [1]; [0], [2], [7]; [1]) \,,  \\
&&~ ([7], [0], [1]; [0], [1], [4]; [1]) \,,~ ([7], [1], [0]; [0], [1], [4]; [1]) \,.~~~
\end{eqnarray}

\begin{table}[h!]
\centering
\resizebox{\textwidth}{!}{
\begin{tabular}{|c|c|c||c|c|c|}
\hline\hline
\multirow{2}{*}{$\ell_L$} &  \multicolumn{2}{c||}{$M_\nu$} & \multirow{2}{*}{$\ell_L$}  & \multicolumn{2}{c|}{$M_\nu$}  \\ \cline{2-3} \cline{5-6}
& $\nu_R\sim ([0],[1],[4])$ & $\nu_R\sim ([0],[2],[7])$ &  & $\nu_R\sim ([0],[1],[4])$ & $\nu_R\sim ([0],[2],[7])$   \\
\hline\hline
$([0], [1], [2])$ & 
$\begin{pmatrix}
0 & \times & 0  \\
\times & 0 & \times  \\
0 & \times & 0  
\end{pmatrix}$ & 
$\begin{pmatrix}
0 & 0 & 0  \\
0 & 0 & \times  \\
0 & \times & \times  
\end{pmatrix}$ & 
$([0], [4], [7])$ & 
$\begin{pmatrix}
0 & 0 & \times  \\
0 & \times & 0  \\
\times & 0 & \times  
\end{pmatrix}$ & 
$\begin{pmatrix}
0 & 0 & 0  \\
0 & \times & \times  \\
0 & \times & 0  
\end{pmatrix}$  \\ \hline 
$([0], [1], [4])$ & 
$\begin{pmatrix}
0 & \times & 0  \\
\times & 0 & 0  \\
0 & 0 & \times  
\end{pmatrix}$ & 
$\begin{pmatrix}
0 & 0 & 0  \\
0 & 0 & \times  \\
0 & \times & \times  
\end{pmatrix}$ & 
$([1], [2], [4])$ & 
$\begin{pmatrix}
0 & \times & 0  \\
\times & 0 & 0  \\
0 & 0 & \times  
\end{pmatrix}$ & 
$\begin{pmatrix}
0 & \times & \times  \\
\times & \times & \times  \\
\times & \times & \times  
\end{pmatrix}$  \\ \hline 
$([0], [1], [7])$ & 
$\begin{pmatrix}
0 & \times & \times  \\
\times & 0 & \times  \\
\times & \times & \times  
\end{pmatrix}$ & 
$\begin{pmatrix}
0 & 0 & 0  \\
0 & 0 & \times  \\
0 & \times & 0  
\end{pmatrix}$ & 
$([1], [2], [7])$ & 
$\begin{pmatrix}
0 & \times & \times  \\
\times & 0 & \times  \\
\times & \times & \times  
\end{pmatrix}$ & 
$\begin{pmatrix}
0 & \times & \times  \\
\times & \times & \times  \\
\times & \times & 0  
\end{pmatrix}$  \\ \hline 
$([0], [2], [4])$ & 
$\begin{pmatrix}
0 & 0 & 0  \\
0 & 0 & 0  \\
0 & 0 & \times  
\end{pmatrix}$ & 
$\begin{pmatrix}
0 & 0 & 0  \\
0 & \times & \times  \\
0 & \times & \times  
\end{pmatrix}$ & 
$([1], [4], [7])$ & 
$\begin{pmatrix}
0 & 0 & \times  \\
0 & \times & 0  \\
\times & 0 & \times  
\end{pmatrix}$ & 
$\begin{pmatrix}
0 & \times & \times  \\
\times & \times & \times  \\
\times & \times & 0  
\end{pmatrix}$  \\ \hline 
$([0], [2], [7])$ & 
$\begin{pmatrix}
0 & 0 & \times  \\
0 & 0 & \times  \\
\times & \times & \times  
\end{pmatrix}$ & 
$\begin{pmatrix}
0 & 0 & 0  \\
0 & \times & \times  \\
0 & \times & 0  
\end{pmatrix}$ & 
$([2], [4], [7])$ & 
$\begin{pmatrix}
0 & 0 & \times  \\
0 & \times & 0  \\
\times & 0 & \times  
\end{pmatrix}$ & 
$\begin{pmatrix}
\times & \times & \times  \\
\times & \times & \times  \\
\times & \times & 0  
\end{pmatrix}$   \\ \hline\hline
\end{tabular}}
\caption{\label{tab:Z13_Seesaw}The neutrino mass matrices $M_\nu$ from $\mathbb{Z}_3$ gauging of $\mathbb{Z}_{13}$ symmetry in type I seesaw, where the Higgs field transforms as $H\sim[1]$. If the assignments for the three generations of left-handed lepton fields are exchanged, the rows and columns of $M_{\nu}$ would be permuted.  }
\end{table}

\subsubsection{$N = 19$}

For $N=19$, $m=7$, then we have $[-1] = [8]$, $[m-1] = [1-2m] = [4]$ and $[m-2] = [5]$. From Eq.~\eqref{eq:C-seesaw-asign}, we know the following assignments of the lepton and Higgs fields can give rise to the neutrino mass matrix of texture $\textbf{C}$,
\begin{eqnarray}
\textbf{C}~:~~ (\ell_L; \nu_R; H)\sim ~& ([8], [4], [5]; [0], [4], [10]; [1])\,,~~
([8], [5], [4]; [0], [4], [10]; [1])\,. \label{eq:C-N19}
\end{eqnarray}
It is worth noting that, for the same assignment of the left-handed leptons as above, the Weinberg operator can yields a neutrino mass matrix of texture \textbf{C}. When only one Higgs field is considered, the analysis presented in section~\ref{sec:weinbop-N19} shows that the charged lepton mass matrix can be diagonal through suitable assignments of the right-handed charged leptons, as indicated in Eq.~\eqref{eq:Yukawa_N19_C}.
For completeness, all textures which can be generated from the $\mathbb{Z}_3$ gauging of $\mathbb{Z}_{19}$ symmetry in type I seesaw mechanism are listed in table~\ref{tab:Z19_Seesaw}. Analogous to the $N=13$ case, besides the assignments in Eq.~\eqref{eq:C-N19}, there are additional assignments of the lepton fields that can produce the texture $\textbf{C}$ for the neutrino mass matrix, as can be seen from table~\ref{tab:Z19_Seesaw}. Nevertheless, these assignments invariably lead to a vanishing mass of the lightest neutrino, which is not allowed for texture $\textbf{C}$.

\renewcommand{\arraystretch}{0.95}
\begin{longtable}{|c||c|c|c|}
\midrule\hline
$M_\nu$	& $\nu_R\sim ([0], [1], [8])$ & $\nu_R\sim ([0], [2], [5])$ & $\nu_R\sim ([0], [4], [10])$   \\
\hline
\endfirsthead
		
\hline
$M_\nu$	& $\nu_R\sim ([0], [1], [8])$ & $\nu_R\sim ([0], [2], [5])$ & $\nu_R\sim ([0], [4], [10])$  \\
\hline
\endhead
\makecell{$\ell_L\sim([0], [1], [2])$} & 
$\begin{pmatrix}
0 & \times & 0  \\
\times & 0 & 0  \\
0 & 0 & 0  
\end{pmatrix}$ & 
$\begin{pmatrix}
0 & 0 & 0  \\
0 & 0 & 0  \\
0 & 0 & 0  
\end{pmatrix}$ & 
$\begin{pmatrix}
0 & 0 & 0  \\
0 & 0 & 0  \\
0 & 0 & \times  
\end{pmatrix}$  \\ \hline 
\makecell{$\ell_L\sim([0], [1], [4])$} & 
$\begin{pmatrix}
0 & \times & 0  \\
\times & 0 & \times  \\
0 & \times & 0  
\end{pmatrix}$ & 
$\begin{pmatrix}
0 & 0 & 0  \\
0 & 0 & \times  \\
0 & \times & 0  
\end{pmatrix}$ & 
$\begin{pmatrix}
0 & 0 & 0  \\
0 & 0 & 0  \\
0 & 0 & 0  
\end{pmatrix}$  \\ \hline 
\makecell{$\ell_L\sim([0], [1], [5])$} & 
$\begin{pmatrix}
0 & \times & \times  \\
\times & 0 & 0  \\
\times & 0 & 0  
\end{pmatrix}$ & 
$\begin{pmatrix}
0 & 0 & 0  \\
0 & 0 & \times  \\
0 & \times & 0  
\end{pmatrix}$ & 
$\begin{pmatrix}
0 & 0 & 0  \\
0 & 0 & 0  \\
0 & 0 & 0  
\end{pmatrix}$  \\ \hline 
\makecell{$\ell_L\sim([0], [1], [8])$} & 
$\begin{pmatrix}
0 & \times & 0  \\
\times & 0 & 0  \\
0 & 0 & \times  
\end{pmatrix}$ & 
$\begin{pmatrix}
0 & 0 & 0  \\
0 & 0 & 0  \\
0 & 0 & \times  
\end{pmatrix}$ & 
$\begin{pmatrix}
0 & 0 & 0  \\
0 & 0 & 0  \\
0 & 0 & \times  
\end{pmatrix}$  \\ \hline 
\makecell{$\ell_L\sim([0], [1], [10])$} & 
$\begin{pmatrix}
0 & \times & 0  \\
\times & 0 & \times  \\
0 & \times & 0  
\end{pmatrix}$ & 
$\begin{pmatrix}
0 & 0 & 0  \\
0 & 0 & \times  \\
0 & \times & \times  
\end{pmatrix}$ & 
$\begin{pmatrix}
0 & 0 & 0  \\
0 & 0 & 0  \\
0 & 0 & 0  
\end{pmatrix}$  \\ \hline 
\makecell{$\ell_L\sim([0], [2], [4])$} & 
$\begin{pmatrix}
0 & 0 & 0  \\
0 & 0 & 0  \\
0 & 0 & 0  
\end{pmatrix}$ & 
$\begin{pmatrix}
0 & 0 & 0  \\
0 & 0 & \times  \\
0 & \times & 0  
\end{pmatrix}$ & 
$\begin{pmatrix}
0 & 0 & 0  \\
0 & \times & \times  \\
0 & \times & 0  
\end{pmatrix}$  \\ \hline 
\makecell{$\ell_L\sim([0], [2], [5])$} & 
$\begin{pmatrix}
0 & 0 & \times  \\
0 & 0 & 0  \\
\times & 0 & 0  
\end{pmatrix}$ & 
$\begin{pmatrix}
0 & 0 & 0  \\
0 & 0 & \times  \\
0 & \times & 0  
\end{pmatrix}$ & 
$\begin{pmatrix}
0 & 0 & 0  \\
0 & \times & \times  \\
0 & \times & 0  
\end{pmatrix}$  \\ \hline 
\makecell{$\ell_L\sim([0], [2], [8])$} & 
$\begin{pmatrix}
0 & 0 & 0  \\
0 & 0 & 0  \\
0 & 0 & \times  
\end{pmatrix}$ & 
$\begin{pmatrix}
0 & 0 & 0  \\
0 & 0 & 0  \\
0 & 0 & \times  
\end{pmatrix}$ & 
$\begin{pmatrix}
0 & 0 & 0  \\
0 & \times & \times  \\
0 & \times & \times  
\end{pmatrix}$  \\ \hline 
\makecell{$\ell_L\sim([0], [2], [10])$} & 
$\begin{pmatrix}
0 & 0 & 0  \\
0 & 0 & 0  \\
0 & 0 & 0  
\end{pmatrix}$ & 
$\begin{pmatrix}
0 & 0 & 0  \\
0 & 0 & \times  \\
0 & \times & \times  
\end{pmatrix}$ & 
$\begin{pmatrix}
0 & 0 & 0  \\
0 & \times & 0  \\
0 & 0 & 0  
\end{pmatrix}$  \\ \hline 
\makecell{$\ell_L\sim([0], [4], [5])$} & 
$\begin{pmatrix}
0 & 0 & \times  \\
0 & 0 & \times  \\
\times & \times & 0  
\end{pmatrix}$ & 
$\begin{pmatrix}
0 & 0 & 0  \\
0 & 0 & 0  \\
0 & 0 & 0  
\end{pmatrix}$ & 
$\begin{pmatrix}
0 & 0 & 0  \\
0 & 0 & \times  \\
0 & \times & 0  
\end{pmatrix}$  \\ \hline 
\makecell{$\ell_L\sim([0], [4], [8])$} & 
$\begin{pmatrix}
0 & 0 & 0  \\
0 & 0 & 0  \\
0 & 0 & \times  
\end{pmatrix}$ & 
$\begin{pmatrix}
0 & 0 & 0  \\
0 & 0 & 0  \\
0 & 0 & \times  
\end{pmatrix}$ & 
$\begin{pmatrix}
0 & 0 & 0  \\
0 & 0 & \times  \\
0 & \times & \times  
\end{pmatrix}$  \\ \hline 
\makecell{$\ell_L\sim([0], [4], [10])$} & 
$\begin{pmatrix}
0 & 0 & 0  \\
0 & 0 & 0  \\
0 & 0 & 0  
\end{pmatrix}$ & 
$\begin{pmatrix}
0 & 0 & 0  \\
0 & 0 & \times  \\
0 & \times & \times  
\end{pmatrix}$ & 
$\begin{pmatrix}
0 & 0 & 0  \\
0 & 0 & 0  \\
0 & 0 & 0  
\end{pmatrix}$  \\ \hline 
\makecell{$\ell_L\sim([0], [5], [8])$} & 
$\begin{pmatrix}
0 & \times & 0  \\
\times & 0 & 0  \\
0 & 0 & \times  
\end{pmatrix}$ & 
$\begin{pmatrix}
0 & 0 & 0  \\
0 & 0 & 0  \\
0 & 0 & \times  
\end{pmatrix}$ & 
$\begin{pmatrix}
0 & 0 & 0  \\
0 & 0 & \times  \\
0 & \times & \times  
\end{pmatrix}$  \\ \hline 
\makecell{$\ell_L\sim([0], [5], [10])$} & 
$\begin{pmatrix}
0 & \times & 0  \\
\times & 0 & \times  \\
0 & \times & 0  
\end{pmatrix}$ & 
$\begin{pmatrix}
0 & 0 & 0  \\
0 & 0 & \times  \\
0 & \times & \times  
\end{pmatrix}$ & 
$\begin{pmatrix}
0 & 0 & 0  \\
0 & 0 & 0  \\
0 & 0 & 0  
\end{pmatrix}$  \\ \hline 
\makecell{$\ell_L\sim([0], [8], [10])$} & 
$\begin{pmatrix}
0 & 0 & 0  \\
0 & \times & 0  \\
0 & 0 & 0  
\end{pmatrix}$ & 
$\begin{pmatrix}
0 & 0 & 0  \\
0 & \times & 0  \\
0 & 0 & \times  
\end{pmatrix}$ & 
$\begin{pmatrix}
0 & 0 & 0  \\
0 & \times & 0  \\
0 & 0 & 0  
\end{pmatrix}$  \\ \hline 
\makecell{$\ell_L\sim([1], [2], [4])$} & 
$\begin{pmatrix}
0 & 0 & \times  \\
0 & 0 & 0  \\
\times & 0 & 0  
\end{pmatrix}$ & 
$\begin{pmatrix}
0 & 0 & \times  \\
0 & 0 & \times  \\
\times & \times & 0  
\end{pmatrix}$ & 
$\begin{pmatrix}
0 & 0 & 0  \\
0 & \times & \times  \\
0 & \times & 0  
\end{pmatrix}$  \\ \hline 
\makecell{$\ell_L\sim([1], [2], [5])$} & 
$\begin{pmatrix}
0 & 0 & 0  \\
0 & 0 & 0  \\
0 & 0 & 0  
\end{pmatrix}$ & 
$\begin{pmatrix}
0 & 0 & \times  \\
0 & 0 & \times  \\
\times & \times & 0  
\end{pmatrix}$ & 
$\begin{pmatrix}
0 & 0 & 0  \\
0 & \times & \times  \\
0 & \times & 0  
\end{pmatrix}$  \\ \hline 
\makecell{$\ell_L\sim([1], [2], [8])$} & 
$\begin{pmatrix}
0 & 0 & 0  \\
0 & 0 & 0  \\
0 & 0 & \times  
\end{pmatrix}$ & 
$\begin{pmatrix}
0 & 0 & 0  \\
0 & 0 & 0  \\
0 & 0 & \times  
\end{pmatrix}$ & 
$\begin{pmatrix}
0 & 0 & 0  \\
0 & \times & \times  \\
0 & \times & \times  
\end{pmatrix}$  \\ \hline 
\makecell{$\ell_L\sim([1], [2], [10])$} & 
$\begin{pmatrix}
0 & 0 & \times  \\
0 & 0 & 0  \\
\times & 0 & 0  
\end{pmatrix}$ & 
$\begin{pmatrix}
0 & 0 & \times  \\
0 & 0 & \times  \\
\times & \times & \times  
\end{pmatrix}$ & 
$\begin{pmatrix}
0 & 0 & 0  \\
0 & \times & 0  \\
0 & 0 & 0  
\end{pmatrix}$  \\ \hline 
\makecell{$\ell_L\sim([1], [4], [5])$} & 
$\begin{pmatrix}
0 & \times & 0  \\
\times & 0 & \times  \\
0 & \times & 0  
\end{pmatrix}$ & 
$\begin{pmatrix}
0 & \times & \times  \\
\times & 0 & 0  \\
\times & 0 & 0  
\end{pmatrix}$ & 
$\begin{pmatrix}
0 & 0 & 0  \\
0 & 0 & \times  \\
0 & \times & 0  
\end{pmatrix}$  \\ \hline 
\makecell{$\ell_L\sim([1], [4], [8])$} & 
$\begin{pmatrix}
0 & \times & 0  \\
\times & 0 & 0  \\
0 & 0 & \times  
\end{pmatrix}$ & 
$\begin{pmatrix}
0 & \times & 0  \\
\times & 0 & 0  \\
0 & 0 & \times  
\end{pmatrix}$ & 
$\begin{pmatrix}
0 & 0 & 0  \\
0 & 0 & \times  \\
0 & \times & \times  
\end{pmatrix}$  \\ \hline 
\makecell{$\ell_L\sim([1], [4], [10])$} & 
$\begin{pmatrix}
0 & \times & \times  \\
\times & 0 & 0  \\
\times & 0 & 0  
\end{pmatrix}$ & 
$\begin{pmatrix}
0 & \times & \times  \\
\times & 0 & \times  \\
\times & \times & \times  
\end{pmatrix}$ & 
$\begin{pmatrix}
0 & 0 & 0  \\
0 & 0 & 0  \\
0 & 0 & 0  
\end{pmatrix}$  \\ \hline 
\makecell{$\ell_L\sim([1], [5], [8])$} & 
$\begin{pmatrix}
0 & 0 & 0  \\
0 & 0 & 0  \\
0 & 0 & \times  
\end{pmatrix}$ & 
$\begin{pmatrix}
0 & \times & 0  \\
\times & 0 & 0  \\
0 & 0 & \times  
\end{pmatrix}$ & 
$\begin{pmatrix}
0 & 0 & 0  \\
0 & 0 & \times  \\
0 & \times & \times  
\end{pmatrix}$  \\ \hline 
\makecell{$\ell_L\sim([1], [5], [10])$} & 
$\begin{pmatrix}
0 & 0 & \times  \\
0 & 0 & \times  \\
\times & \times & 0  
\end{pmatrix}$ & 
$\begin{pmatrix}
0 & \times & \times  \\
\times & 0 & \times  \\
\times & \times & \times  
\end{pmatrix}$ & 
$\begin{pmatrix}
0 & 0 & 0  \\
0 & 0 & 0  \\
0 & 0 & 0  
\end{pmatrix}$  \\ \hline 
\makecell{$\ell_L\sim([1], [8], [10])$} & 
$\begin{pmatrix}
0 & 0 & \times  \\
0 & \times & 0  \\
\times & 0 & 0  
\end{pmatrix}$ & 
$\begin{pmatrix}
0 & 0 & \times  \\
0 & \times & 0  \\
\times & 0 & \times  
\end{pmatrix}$ & 
$\begin{pmatrix}
0 & 0 & 0  \\
0 & \times & 0  \\
0 & 0 & 0  
\end{pmatrix}$  \\ \hline 
\makecell{$\ell_L\sim([2], [4], [5])$} & 
$\begin{pmatrix}
0 & 0 & 0  \\
0 & 0 & \times  \\
0 & \times & 0  
\end{pmatrix}$ & 
$\begin{pmatrix}
0 & \times & \times  \\
\times & 0 & 0  \\
\times & 0 & 0  
\end{pmatrix}$ & 
$\begin{pmatrix}
\times & \times & \times  \\
\times & 0 & \times  \\
\times & \times & 0  
\end{pmatrix}$  \\ \hline 
\makecell{$\ell_L\sim([2], [4], [8])$} & 
$\begin{pmatrix}
0 & 0 & 0  \\
0 & 0 & 0  \\
0 & 0 & \times  
\end{pmatrix}$ & 
$\begin{pmatrix}
0 & \times & 0  \\
\times & 0 & 0  \\
0 & 0 & \times  
\end{pmatrix}$ & 
$\begin{pmatrix}
\times & \times & \times  \\
\times & 0 & \times  \\
\times & \times & \times  
\end{pmatrix}$  \\ \hline 
\makecell{$\ell_L\sim([2], [4], [10])$} & 
$\begin{pmatrix}
0 & 0 & 0  \\
0 & 0 & 0  \\
0 & 0 & 0  
\end{pmatrix}$ & 
$\begin{pmatrix}
0 & \times & \times  \\
\times & 0 & \times  \\
\times & \times & \times  
\end{pmatrix}$ & 
$\begin{pmatrix}
\times & \times & 0  \\
\times & 0 & 0  \\
0 & 0 & 0  
\end{pmatrix}$  \\ \hline 
\makecell{$\ell_L\sim([2], [5], [8])$} & 
$\begin{pmatrix}
0 & 0 & 0  \\
0 & 0 & 0  \\
0 & 0 & \times  
\end{pmatrix}$ & 
$\begin{pmatrix}
0 & \times & 0  \\
\times & 0 & 0  \\
0 & 0 & \times  
\end{pmatrix}$ & 
$\begin{pmatrix}
\times & \times & \times  \\
\times & 0 & \times  \\
\times & \times & \times  
\end{pmatrix}$  \\ \hline 
\makecell{$\ell_L\sim([2], [5], [10])$} & 
$\begin{pmatrix}
0 & 0 & 0  \\
0 & 0 & \times  \\
0 & \times & 0  
\end{pmatrix}$ & 
$\begin{pmatrix}
0 & \times & \times  \\
\times & 0 & \times  \\
\times & \times & \times  
\end{pmatrix}$ & 
$\begin{pmatrix}
\times & \times & 0  \\
\times & 0 & 0  \\
0 & 0 & 0  
\end{pmatrix}$  \\ \hline 
\makecell{$\ell_L\sim([2], [8], [10])$} & 
$\begin{pmatrix}
0 & 0 & 0  \\
0 & \times & 0  \\
0 & 0 & 0  
\end{pmatrix}$ & 
$\begin{pmatrix}
0 & 0 & \times  \\
0 & \times & 0  \\
\times & 0 & \times  
\end{pmatrix}$ & 
$\begin{pmatrix}
\times & \times & 0  \\
\times & \times & 0  \\
0 & 0 & 0  
\end{pmatrix}$  \\ \hline 
\makecell{$\ell_L\sim([4], [5], [8])$} & 
$\begin{pmatrix}
0 & \times & 0  \\
\times & 0 & 0  \\
0 & 0 & \times  
\end{pmatrix}$ & 
$\begin{pmatrix}
0 & 0 & 0  \\
0 & 0 & 0  \\
0 & 0 & \times  
\end{pmatrix}$ & 
$\begin{pmatrix}
0 & \times & \times  \\
\times & 0 & \times  \\
\times & \times & \times  
\end{pmatrix}$  \\ \hline 
\makecell{$\ell_L\sim([4], [5], [10])$} & 
$\begin{pmatrix}
0 & \times & 0  \\
\times & 0 & \times  \\
0 & \times & 0  
\end{pmatrix}$ & 
$\begin{pmatrix}
0 & 0 & \times  \\
0 & 0 & \times  \\
\times & \times & \times  
\end{pmatrix}$ & 
$\begin{pmatrix}
0 & \times & 0  \\
\times & 0 & 0  \\
0 & 0 & 0  
\end{pmatrix}$  \\ \hline 
\makecell{$\ell_L\sim([4], [8], [10])$} & 
$\begin{pmatrix}
0 & 0 & 0  \\
0 & \times & 0  \\
0 & 0 & 0  
\end{pmatrix}$ & 
$\begin{pmatrix}
0 & 0 & \times  \\
0 & \times & 0  \\
\times & 0 & \times  
\end{pmatrix}$ & 
$\begin{pmatrix}
0 & \times & 0  \\
\times & \times & 0  \\
0 & 0 & 0  
\end{pmatrix}$  \\ \hline 
\makecell{$\ell_L\sim([5], [8], [10])$} & 
$\begin{pmatrix}
0 & 0 & \times  \\
0 & \times & 0  \\
\times & 0 & 0  
\end{pmatrix}$ & 
$\begin{pmatrix}
0 & 0 & \times  \\
0 & \times & 0  \\
\times & 0 & \times  
\end{pmatrix}$ & 
$\begin{pmatrix}
0 & \times & 0  \\
\times & \times & 0  \\
0 & 0 & 0  
\end{pmatrix}$  \\ \hline
\midrule
\caption{\label{tab:Z19_Seesaw}The neutrino mass matrices $M_\nu$ from $\mathbb{Z}_3$ gauging of $\mathbb{Z}_{19}$ symmetry in type-I seesaw, where the Higgs field transforms as $H\sim[1]$. }
\end{longtable}

\section{General $\mathbb{Z}_n$ gauging of $\mathbb{Z}_N$ symmetry\label{sec:Zn_gauging_ZN}}

Now we generalize $\mathbb{Z}_3$ gauging of $\mathbb{Z}_N$ symmetry to more general $\mathbb{Z}_n$ gauging of $\mathbb{Z}_N$ symmetry, whose generator is denoted as $a$ which satisfies $a^N = e$. The $\mathbb{Z}_n$ symmetry is a subgroup of the automorphism group of $\mathbb{Z}_N$, and it is generated by the element $b$ which fulfills
\begin{eqnarray}
b^{-1} a b = a^m \,,~~~b^n = e\,,
\end{eqnarray}
where $m$ is an integer. Similar to the discussion in section~\ref{sec:Z3_gauging_ZN}, the parameter $m$ must fulfill 
\begin{eqnarray}
m^n - 1 \equiv 0 ~ ({\rm mod}~ N) ~~~\text{and}~~~m^d - 1 \not\equiv 0 ~({\rm mod} ~ N)\,, \label{eq:cond-nGaugeN}
\end{eqnarray} 
for any proper divisor $d$ of $n$. Otherwise, the theory would reduce to a $\mathbb{Z}_d$ gauging of $\mathbb{Z}_N$ symmetry. We shall not consider the trivial solution $m=1$ in the following. In this scenario, the equivalence class of $\mathbb{Z}_N$ is given by
\begin{eqnarray}
[k] = \left\{ a^{k}, a^{km}, \cdots , a^{km^{n-1}} \right\} = \left\{ \left.a^{k m^t} \right| 0 \leq t < n\right\}\,,
\end{eqnarray}
and the fusion rule reads as
\begin{eqnarray}
[k] \times [l] = [k + l] + [k + l m] + \cdots + [k + l m^{n-1}] = \sum_{t=0}^{n-1} [k + l m^{t}] \,.
\end{eqnarray}

For $n=2$, the parameter $N$ can be any positive integer larger than two, and all possible gauging schemes are discussed in appendix~\ref{sec:Z2_gauging_of_ZN}. When $n > 2$, only certain values of $N$ and $m$ can satisfy the above conditions. For each given integer $n$, there exists a systematic procedure to identify the values of $N$ for which the above conditions in Eq.~\eqref{eq:cond-nGaugeN} hold. To solve the above equations, the problem reduces to identifying a $\mathbb{Z}_n$ subgroup of the automorphism group of $\mathbb{Z}_N$, denoted by ${\rm Aut}(\mathbb{Z}_N)$. It is well known that ${\rm Aut}(\mathbb{Z}_N)$ is Abelian. Determining the solution for $m$ is therefore amounts to finding a generator of such a $\mathbb{Z}_n$ subgroup within ${\rm Aut}(\mathbb{Z}_N)$. Generally $N$ can be factored as $N=p_1^{\alpha_1} \cdots p_s^{\alpha_s}$ with each $p_i$ a prime number and $\alpha_i$ being positive integer, then by Chinese remainder theorem, the automorphism group ${\rm Aut}(\mathbb{Z}_N)$ is isomorphic to the direct product of ${\rm Aut}(\mathbb{Z}_{p_i^{\alpha_i}})$, i.e.
\begin{eqnarray}
{\rm Aut}(\mathbb{Z}_N) \cong {\rm Aut}(\mathbb{Z}_{p_1^{\alpha_1}}) \times \cdots \times {\rm Aut}(\mathbb{Z}_{p_s^{\alpha_s}}) \,.
\end{eqnarray}
The automorphism group ${\rm Aut}(\mathbb{Z}_{p^{\alpha}}) $ is a cyclic group for $p>2$ or $p=2$, $\alpha\leq 2$ with~\cite{gauss1870disquisitiones}
\begin{eqnarray}
{\rm Aut}(\mathbb{Z}_{p^\alpha}) \cong \mathbb{Z}_{(p-1)p^{\alpha-1}} \,. \label{eq:AuT-Zpalpha}
\end{eqnarray}
When $p=2$ and $\alpha\geq 3$, it is isomorphic to
\begin{eqnarray}\label{eq:Aut_2alpha}
{\rm Aut}(\mathbb{Z}_{2^\alpha}) \cong \mathbb{Z}_2 \times \mathbb{Z}_{2^{\alpha-2}} \,,
\end{eqnarray}
where the group $\mathbb{Z}_2 \times \mathbb{Z}_{2^{\alpha-2}}$ can be generated by $-1$ and $5$ respectively. Meanwhile, the cyclic group $\mathbb{Z}_{(p-1)p^{\alpha-1}}$ in Eq.~\eqref{eq:AuT-Zpalpha} can be generated by a number known as the primitive root of  $p^\alpha$. Here we denote the primitive root as $g$ which satisfies
\begin{eqnarray}\label{eq:primitive_root}
g^{(p-1)p^{\alpha-1}} \equiv 1 ~ ({\rm mod} ~ p^\alpha) ~~~ \text{ and } ~~~ 
g^d \not\equiv 1 ~ ({\rm mod} ~ p^\alpha) ~~~ \text{ for }~~~\forall d<(p-1)p^{\alpha-1} \,.
\end{eqnarray}
The structure of the automorphism group ${\rm Aut}(\mathbb{Z}_N)$ implies that it is sufficient to analyze the basic building block, namely the $\mathbb{Z}_{n}$ gauging of $\mathbb{Z}_{N}$ with $N=p^\alpha$ and $n=q^\beta$, where both $p$ and $q$ are prime numbers. Requiring $\mathbb{Z}_n$ to be embedded as a subgroup of ${\rm Aut}(\mathbb{Z}_N)$ yields the divisibility condition $n | p^{\alpha-1} (p-1)$ for $p\geq 3$, or $n | 2^{\alpha-2}$ for $p=2$ and $\alpha\geq3$.
As a consequence, if $q=2$ and $\beta > 1$, then $N$ must be divisible either by $2^{\beta+2}$ or by a prime number $p$ of the form $p=1\;  (\text{mod}\; n)$.
If $q>2$, then $N$ must be divisible either by $q^{\beta+1}$ or be divisible by a prime number $p$ satisfying $p=1\;(\text{mod}\; n)$. 
In this case, we can take $m=g^{p^{\alpha-1}(p-1)/q^\beta}$ as our solution to $m^{q^\beta} \equiv 1$ modulo $p^\alpha$.

If $N$ can be factorized into two coprime numbers $N_1$ and $N_2$ as $N=N_1 N_2$, then the Eq.~\eqref{eq:cond-nGaugeN} can be decomposed as
\begin{eqnarray}
m^{n} \equiv 1 ~({\rm mod}~N_1)\,, ~~~ m^{n} \equiv 1 ~({\rm mod}~N_2) \,,
\end{eqnarray}
However, $n$ is usually not the least positive number such that $m^{n} \equiv 1 ~({\rm mod}~N_i)$. We denote $n_1$ and $n_2$ to be the least positive number satisfying 
\begin{eqnarray}\label{eq:cond-nGaugeN_dec}
m^{n_1} \equiv 1 ~({\rm mod}~N_1)\,, ~~~ m^{n_2} \equiv 1 ~({\rm mod}~N_2) \,,
\end{eqnarray}
respectively. Then we have $n = {\rm lcm}(n_1, n_2)$ which is the least common multiple of the numbers $n_1$ and $n_2$~\footnote{Let $n' = {\rm lcm}(n_1, n_2)$. If $n' < n$, then 
\begin{eqnarray}
m^{n'} \equiv 1 ~({\rm mod}~N_1)\,, ~~~ m^{n'} \equiv 1 ~({\rm mod}~N_2) \,,
\end{eqnarray}
which implies $m^{n'} \equiv 1 ~({\rm mod}~N)$ in contradiction with $m^{n} \equiv 1 ~({\rm mod}~N)$. If $n'>n$, then $n_1|n$ and $n_2|n$ cannot be fulfilled simultaneously. Without loss of generality, we take $n_1\not| n$ and $n=c n_1 + d$ with $1\leq d < n_1$. Therefore
\begin{eqnarray}
1 \equiv m^n = m^{cn_1 + d} \equiv m^d  ~({\rm mod}~N_1)\,,
\end{eqnarray}
which contradicts to the fact that $n_1$ is the least one. Therefore, $n=n' = {\rm lcm}(n_1, n_2)$.}. Specially, if $n_1$ and $n_2$ are coprime numbers, then $n=n_1 n_2$.  
In this case, we can use the Chinese remainder theorem to write the equivalence class of $\mathbb{Z}_N$ as 
\begin{eqnarray}
[k] &=& \left\{ \left. a^{k m^{r x_0 + s y_0}} \right| 0\leq r < n_1, 0\leq s < n_2 \right\}  
\end{eqnarray}
where the integers $x_0$ and $y_0$ satisfy
\begin{eqnarray}
x_0 \equiv 1 ~({\rm mod}~n_1)\,, ~~x_0 \equiv 0 ~({\rm mod}~n_2)\,, ~~ y_0 \equiv 1 ~({\rm mod}~n_2)\,, ~~y_0 \equiv 0 ~({\rm mod}~n_1) \,.
\end{eqnarray} 
We can see that 
\begin{eqnarray}
k m^{r x_0 + s y_0} \equiv k m^r ~ ({\rm mod}~N_1) ~~~ \text{ and } ~~~ k m^{r x_0 + s y_0} \equiv k m^s ~ ({\rm mod}~N_2) \,.
\end{eqnarray}
Therefore, the equivalence class of $\mathbb{Z}_N$ can be regarded as the product of the equivalence classes of $\mathbb{Z}_{N_1}$ and $\mathbb{Z}_{N_2}$, i.e.
\begin{eqnarray}
[k]_N = ([k]_{N_1}, [k]_{N_2}) \,.
\end{eqnarray}
where the subscript serves to distinguish the equivalence classes of the $\mathbb{Z}_N$, $\mathbb{Z}_{N_1}$ and $\mathbb{Z}_{N_2}$ under consideration.
It is straightforward to check that the fusion rule fulfills the following identity  
\begin{eqnarray}\label{eq:fusion_rule_product}
[k]_{N} \times [l]_{N} &=& \sum_{r=0}^{n_1-1} \sum_{s=0}^{n_2-1} ([k + l m^r]_{N_1}, [k + l m^s]_{N_2}) = ([k]_{N_1} \times [l]_{N_1}, [k]_{N_2} \times [l]_{N_2})  \,.
\end{eqnarray}
Hence the $\mathbb{Z}_n$ gauging of $\mathbb{Z}_N$ can be regarded as the product of $\mathbb{Z}_{n_1}$ gauging of $\mathbb{Z}_{N_1}$ and $\mathbb{Z}_{n_2}$ gauging of $\mathbb{Z}_{N_2}$.

Let us give a few examples for relatively small values of $n$.
For $n=4$, we need to identify the values of $N$ for which ${\rm Aut}(\mathbb{Z}_N)$ contains $\mathbb{Z}_4$ subgroup.
This is equivalent to finding the combination of $(N, m)$ satisfying 
\begin{eqnarray}\label{eq:n4_modEq}
m^4 \equiv 1 ~ ({\rm mod}~N) ~~~\text{ and } ~~~m^2\not\equiv 1 ~ ({\rm mod}~N) \,. 
\end{eqnarray}
If $N$ is a prime number, one needs to solve the equation
\begin{eqnarray}
m^2 + 1 \equiv 0 ~ ({\rm mod}~N)  \,.
\end{eqnarray}
Since $-1$ is not the quadratic residue modulo $N$ if the prime number $N\equiv 3 ~({\rm mod}~4)$~\cite{enwiki:1335677700}, therefore $N$ must be a prime number of the form $4k+1$, i.e. $N\equiv 1 ~ ({\rm mod} ~ 4)$. The possible combinations of $(N,m)$ for prime number $N$ are
\begin{eqnarray}
(N,m) = (5, 2)\,,~ (13, 5)\,,~ (17, 4)\,,~(29, 12)\,,\cdots \,.
\end{eqnarray}
If $N$ is not a prime number, the possible combinations can be
\begin{eqnarray}
(N,m) = (10, 3)\,,~ (15,2)\,,~ (15,7)\,,~ (16,3)\,,~(16,5) \,,\cdots \,.
\end{eqnarray}
For a given $N$, there are often multiple values of $m$ that serve as solutions of Eq.~\eqref{eq:n4_modEq}. Here, if different values of $m$ correspond to the same equivalence class, we list only the smallest representative. Whenever multiple values of $m$ are presented for a given $N$, they correspond to distinct equivalence classes.

For $n=5$, we need to find out the values of $N$ so that $\mathbb{Z}_5$ is the subgroup of ${\rm Aut}(\mathbb{Z}_N)$. The corresponding $m$ have to satisfy 
\begin{eqnarray}
m^5 \equiv 1 ~ ({\rm mod}~N)  \,.
\end{eqnarray}
According to the general results below Eq.~\eqref{eq:primitive_root}, we know the value $N$ should be divisible by a prime number of the form $5k+1$ or be divisible by $25$.
If $N$ is a prime number, the possible combinations of $(N,m)$ are
\begin{eqnarray}
(N,m) = (11, 3) \,,~ (31, 2) \,,~ (41, 10) \,,~ (61,9) \,,\cdots \,.
\end{eqnarray}
If $N$ is not a prime number, we have
\begin{eqnarray}
(N, m) = (22, 3) \,,~ (25, 6) \,,~ (33, 4) \,,~ (44,5) \,, \cdots\,.
\end{eqnarray}
For $n=6$, $N$ should be divisible by $9$ or a prime number of the form $3k+1$. Each value of $N$ that allows a $\mathbb{Z}_3$ gauging of $\mathbb{Z}_N$ can also accommodate a $\mathbb{Z}_6$ gauging of $\mathbb{Z}_N$. Nevertheless, different values of $m$ are required. In the case $n=6$, the parameter $m$ must fulfill the following conditions
\begin{eqnarray}
m^6 \equiv 1  ~ ({\rm mod}~N) ~~~ \text{ and }~~~ m^3 \not\equiv 1  ~ ({\rm mod}~N) ~~~ \text{ and }~~~ m^2 \not\equiv 1  ~ ({\rm mod}~N)\,.
\end{eqnarray}
If $N$ is a prime number, we find that the following values of $m$ and $N$ are allowed, 
\begin{eqnarray}
(N, m) = (7, 3) \,,~ (13, 4) \,,~ (19, 8) \,,~ (31, 6) \,,\cdots \,.
\end{eqnarray}
If $N$ is not a prime number, we have 
\begin{eqnarray}
\nonumber (N, m) &=& (9, 2) \,,~ (14, 3) \,,~ (18, 5) \,,~ (21,2) \,,~ (21, 5)\,,~ (21,10)\,,~ (26,17)\,,~ (27,8)\,,~ \\
&&  (28,3) \,,~ (28,5)\,,~ (28,11)\,,~ (35,4) \,,~(35,19)\,,~ (35,26)  \cdots \,.
\end{eqnarray}
Here we give an example for $\mathbb{Z}_6$ gauging of $\mathbb{Z}_{35}$.  
In the present example, $N=35$ can be written as a product of two coprime integers, $N = N_1 N_2$, with $N_1=5$ and $N_2=7$. The corresponding values of $(n_1, n_2)$ can be $(2,3)$, $(1,6)$ and $(2,6)$. The automorphism group of $\mathbb{Z}_{35}$ is ${\rm Aut}(\mathbb{Z}_{35}) \cong \mathbb{Z}_4 \times \mathbb{Z}_{6}$, which has three different $\mathbb{Z}_6$ subgroups.
Therefore, we obtain three distinct values of $m$, namely $m=4$, $m=19$, and $m=26$, which correspond to three different realizations of $\mathbb{Z}_6$ gauging of $\mathbb{Z}_{35}$.
When $n_1 = 2$ and $n_2 = 3$, we have 
$m=4$ satisfying Eq.~\eqref{eq:cond-nGaugeN_dec}.
In this realization, the equivalence classes of $\mathbb{Z}_{35}$ are given by 
\begin{eqnarray}
\nonumber [0]_{35} &=& \{ e \} = ( [0]_5, [0]_7 ) \,,~~~ [1]_{35} = \{ a, a^4, a^9, a^{11}, a^{16}, a^{29} \} = ( [1]_5, [1]_7 ) \,,  \\
\nonumber [2]_{35} &=& \{ a^2, a^8, a^{18}, a^{22}, a^{23}, a^{32} \} = ( [2]_5, [1]_7 ) \,,~~~  [3]_{35} =  \{ a^3, a^{12}, a^{13}, a^{17}, a^{27}, a^{33} \} = ( [2]_5, [3]_7 ) \,, \\
\nonumber [5]_{35} &=& \{ a^5, a^{10}, a^{20} \} = ( [0]_5, [3]_7 )  \,,~~~  [6]_{35} = \{ a^6, a^{19}, a^{24}, a^{26}, a^{31}, a^{34} \} = ( [1]_5, [3]_7) \,, \\
\nonumber [7]_{35} &=& \{ a^7, a^{28} \} = ( [2]_5, [0]_7 ) \,,~~~  [14]_{35} = \{ a^{14}, a^{21} \} = ( [1]_5, [0]_7 ) \,, \\
~ [15]_{35} &=& \{ a^{15}, a^{25}, a^{30} \} = ( [0]_5, [1]_7 ) \,,
\end{eqnarray}
The equivalence classes in $\mathbb{Z}_6$ gauging of $\mathbb{Z}_{35}$ are the product of these of $\mathbb{Z}_{2}$ gauging of $\mathbb{Z}_5$ and $\mathbb{Z}_3$ gauging of $\mathbb{Z}_7$. The fusion rules can be obtained as the tensor product of the fusion rules of the $\mathbb{Z}_2$ gauging of $\mathbb{Z}_5$ and the $\mathbb{Z}_3$ gauging of $\mathbb{Z}_7$, as given in  Eq.~\eqref{eq:fusion_rule_product}.
\begin{eqnarray}
\nonumber [1]_{35} \times [1]_{35} &=& [2]_{35} + 2[3]_{35} + 4[5]_{35} + 2[15]_{35}  \,, ~~[1]_{35} \times [2]_{35} = [1]_{35} + [2]_{35} + 2[3]_{35} + 2[6]_{35}  \,,\\ 
\nonumber [1]_{35} \times [3]_{35} &=& [1]_{35} + [2]_{35} + [3]_{35} + [6]_{35} + 3[7]_{35} + 3[14]_{35}  \,, ~~[1]_{35} \times [5]_{35} = [1]_{35} + [6]_{35} + 3[14]_{35}\\ 
\nonumber [1]_{35} \times [6]_{35} &=& 6[0]_{35} + [2]_{35} + [3]_{35} + 2[5]_{35} + 3[7]_{35} + 2[15]_{35}  \,, ~~~[1]_{35} \times [7]_{35} = [1]_{35} + [2]_{35}  \,,\\ 
\nonumber [1]_{35} \times [14]_{35} &=& [2]_{35} + 2[15]_{35}  \,, ~~~[1]_{35} \times [15]_{35} = [1]_{35} + 2[6]_{35}  \,,\\ 
\nonumber [2]_{35} \times [2]_{35} &=& [1]_{35} + 4[5]_{35} + 2[6]_{35} + 2[15]_{35}  \,, ~~~\\ 
\nonumber [2]_{35} \times [3]_{35} &=& 6[0]_{35} + [1]_{35} + 2[5]_{35} + [6]_{35} + 3[14]_{35} + 2[15]_{35}  \,, ~~[2]_{35} \times [5]_{35} = [2]_{35} + [3]_{35} + 3[7]_{35}  \,,\\ 
\nonumber [2]_{35} \times [6]_{35} &=& [1]_{35} + [2]_{35} + [3]_{35} + [6]_{35} + 3[7]_{35} + 3[14]_{35}  \,, ~~~[2]_{35} \times [7]_{35} = [1]_{35} + 2[15]_{35}  \,,\\ 
\nonumber [2]_{35} \times [14]_{35} &=& [1]_{35} + [2]_{35}  \,, ~~~[2]_{35} \times [15]_{35} = [2]_{35} + 2[3]_{35}  \,,\\ 
\nonumber [3]_{35} \times [3]_{35} &=& 2[1]_{35} + 2[5]_{35} + [6]_{35} + 4[15]_{35}  \,, ~~~[3]_{35} \times [5]_{35} = 2[2]_{35} + [3]_{35}  \,,\\ 
\nonumber [3]_{35} \times [6]_{35} &=& 2[1]_{35} + 2[2]_{35} + [3]_{35} + [6]_{35}  \,, ~~~[3]_{35} \times [7]_{35} = 2[5]_{35} + [6]_{35}  \,,\\ 
\nonumber [3]_{35} \times [14]_{35} &=& [3]_{35} + [6]_{35}  \,, ~~~[3]_{35} \times [15]_{35} = [2]_{35} + [3]_{35} + 3[7]_{35}  \,,\\ 
\nonumber [5]_{35} \times [5]_{35} &=& [5]_{35} + 2[15]_{35}  \,, ~~~[5]_{35} \times [6]_{35} = 2[1]_{35} + [6]_{35}  \,,[5]_{35} \times [7]_{35} = [3]_{35}  \,,\\ 
\nonumber [5]_{35} \times [14]_{35} &=& [6]_{35}  \,, ~~~[5]_{35} \times [15]_{35} = 3[0]_{35} + [5]_{35} + [15]_{35}  \,,\\ 
\nonumber [6]_{35} \times [6]_{35} &=& 2[2]_{35} + [3]_{35} + 2[5]_{35} + 4[15]_{35}  \,, ~~~[6]_{35} \times [7]_{35} = [3]_{35} + [6]_{35}  \,,\\ 
\nonumber [6]_{35} \times [14]_{35} &=& [3]_{35} + 2[5]_{35}  \,, ~~~[6]_{35} \times [15]_{35} = [1]_{35} + [6]_{35} + 3[14]_{35}  \,,\\ 
\nonumber [7]_{35} \times [7]_{35} &=& 2[0]_{35} + [14]_{35}  \,, ~~~[7]_{35} \times [14]_{35} = [7]_{35} + [14]_{35}  \,,[7]_{35} \times [15]_{35} = [2]_{35}  \,,\\ 
~ [14]_{35} \times [14]_{35} &=& 2[0]_{35} + [7]_{35}  \,, ~~~[14]_{35} \times [15]_{35} = [1]_{35}  \,,[15]_{35} \times [15]_{35} = 2[5]_{35} + [15]_{35}  \,. \label{eq:Z6-Z35-fusion}
\end{eqnarray}
We see $[0]_{35} \subset ([k_1]_5, [l_1]_7) \times ([k_2]_5, [l_2]_7)$ if and only if $[0]_5 \subset [k_1]_5 \times [k_2]_5$ and $[0]_7 \subset [l_1]_7 \times [l_2]_7$.

When $n_1 = 1$ and $n_2 = 6$, we have $m=26$. The equivalence classes are
\begin{eqnarray}
\nonumber [0]_{35} &=& \{ e \} = ( [0]_5, [0]_7 ) \,,~~~ [1]_{35} = \{ a, a^6, a^{11}, a^{16}, a^{26}, a^{31} \} = ( [1]_5, [1]_7 ) \,,\\
\nonumber [2]_{35} &=& \{ a^2, a^{12}, a^{17}, a^{22}, a^{27}, a^{32} \} = ( [2]_5, [1]_7 ) \,,~~~ [3]_{35} = \{ a^3, a^8, a^{13}, a^{18}, a^{23}, a^{33} \} = ([3]_5, [1]_7)\,,\\
\nonumber [4]_{35} &=& \{ a^4, a^9, a^{19}, a^{24}, a^{29}, a^{34} \} = ([4]_5, [1]_7) \,,~~~ [5]_{35} = \{ a^5, a^{10}, a^{15}, a^{20}, a^{25}, a^{30} \} = ([0]_5, [1]_7) \,,  \\
\nonumber [7]_{35} &=& \{ a^7 \} = ([2]_5, [0]_7) \,,~~~ [14]_{35} = \{ a^{14} \} = ([4]_5, [0]_7) \,,  \\
~ [21]_{35} &=& \{ a^{21} \} = ([1]_5, [0]_7) \,,~~~ [28]_{35} = \{ a^{28} \} = ([3]_5, [0]_7) \,.
\end{eqnarray}
In this case, every equivalence class of $\mathbb{Z}_5$ coincides with the group element of $\mathbb{Z}_5$ due to $n_1=1$. This fusion rule can be obtained by product of $\mathbb{Z}_5$ group multiplication rule and $\mathbb{Z}_6$ gauging of $\mathbb{Z}_7$ non-invertible fusion rule.

When $n_1 = 2$ and $n_2=6$, we have $m=19$. The equivalence classes are
\begin{eqnarray}
\nonumber [0]_{35} &=& \{ e \} \,,~~~  [1]_{35} = \{ a, a^{11}, a^{16}, a^{19}, a^{24}, a^{34} \} \,,~~~ [2]_{35} = \{ a^2, a^3, a^{13}, a^{22}, a^{32}, a^{33} \}\,, \\
\nonumber [4]_{35} &=& \{ a^4, a^6, a^9, a^{26}, a^{29}, a^{31} \} \,,~~~ [5]_{35} = \{ a^5, a^{10}, a^{15}, a^{20}, a^{25}, a^{30} \} \,,~~~ [7]_{35} = \{ a^7, a^{28} \} \,, \\
~ [8]_{35} &=& \{ a^8, a^{12}, a^{17}, a^{18}, a^{23}, a^{27} \} \,,~~~ [14]_{35} = \{ a^{14}, a^{21} \} \,.
\end{eqnarray}
These equivalence classes cannot be expressed as the product of equivalence classes of $\mathbb{Z}_5$ and the equivalence classes of $\mathbb{Z}_7$ because $n_1=2$ and $n_2=6$ are not coprime numbers. 

\subsection{Applying the fusion rules of $\mathbb{Z}_6$ gauging of $\mathbb{Z}_{35}$ symmetry with $m=4$ to Weinberg operator  }

In this subsection, we give an explicit example by applying the fusion algebra of $\mathbb{Z}_6$ gauging of $\mathbb{Z}_{35}$ symmetry with $m=4$ to the Weinberg operator. The lepton doublet and Higgs field are assigned to the equivalency classes $\ell_L \sim ([0]_{35}, [1]_{35}, [3]_{35})$ and $H\sim [1]_{35}$. Using the fusion rules in Eq.~\eqref{eq:Z6-Z35-fusion}, we find that the neutrino mass matrix is of the following pattern 
\begin{eqnarray}
M_\nu = \begin{pmatrix}
0 & 0 & \times  \\
0 & \times & \times  \\
\times & \times & \times 
\end{pmatrix} \,, \label{eq:mnu-Z6-Z35}
\end{eqnarray}
which is exactly the texture $\textbf{A}_1$. Because the $\mathbb{Z}_6$ gauging of $\mathbb{Z}_{35}$ is isomorphic to the direct product of the $\mathbb{Z}_2$ gauging of $\mathbb{Z}_5$ and the $\mathbb{Z}_3$ gauging of $\mathbb{Z}_7$, with $[0]_{35}=([0]_5,[0]_7)$, $[1]_{35}=([1]_5,[1]_7)$, and $[3]_{35}=([2]_5,[3]_7)$, the same neutrino mass matrix in Eq.~\eqref{eq:mnu-Z6-Z35} can be reproduced from the $\mathbb{Z}_2$ gauging of $\mathbb{Z}_5$ together with the $\mathbb{Z}_3$ gauging of $\mathbb{Z}_7$, as discussed in~\cite{Dong:2025jra}. Under the $\mathbb{Z}_2$ gauging of $\mathbb{Z}_5$ symmetry, the lepton doublets and Higgs fields are labeled by $\ell_L\sim ( [0]_5, [1]_5, [2]_5 )$, $H\sim[1]_5$,  then the neutrino mass matrix derived from the Weinberg operator takes the following form~\cite{Kobayashi:2025ldi}:
\begin{eqnarray}\label{eq:Z2_Z5_texture}
M_\nu = \begin{pmatrix}
\times & 0 & \times  \\
0 & \times & \times  \\
\times & \times & \times 
\end{pmatrix} \,.
\end{eqnarray}  
Under the $\mathbb{Z}_3$ gauging of $\mathbb{Z}_7$, we can assign $\ell_L \sim ([0]_7, [1]_7, [3]_7)$ and $H\sim [1]_7$, which leads to the following neutrino mass texture
\begin{eqnarray}\label{eq:Z3_Z7_texture}
M_\nu = \begin{pmatrix}
0 & \times & \times  \\
\times & \times & \times  \\
\times & \times & \times 
\end{pmatrix} \,,
\end{eqnarray}
which coincides with the result obtained in Eq.~\eqref{eq:Weinberg_N7}. The light neutrino mass matrix in Eq.~\eqref{eq:mnu-Z6-Z35} can be obtained by combining the textures in Eq~\eqref{eq:Z2_Z5_texture} and Eq.~\eqref{eq:Z3_Z7_texture}. In short, the selection rules from the $\mathbb{Z}_2$ gauging of $\mathbb{Z}_5$ and the $\mathbb{Z}_3$ gauging of $\mathbb{Z}_7$ can be combined via a tensor product. The resulting fusion algebra is equivalent to that of $\mathbb{Z}_6$ gauging of $\mathbb{Z}_{35}$ with $m = 4$.

\section{Conclusion\label{sec:conclusion}}

In an effort to reduce the number of free parameters, the texture-zero ansatz has been widely investigated. In the physical basis where the charged leptons are diagonal, Majorana neutrino mass matrices may possess no more than two vanishing entries in order to remain consistent with experimental data. This requirement yields seven inequivalent textures, labeled $\mathbf{A}_{1,2}$, $\mathbf{B}_{1,2,3,4}$, and $\mathbf{C}$.

It has recently been demonstrated that fusion rules stemming from non-invertible symmetries can impose nontrivial constraints on the structure of quark and lepton mass matrices, thereby enforcing specific patterns of texture zeros~\cite{Kobayashi:2024cvp,Kobayashi:2025znw,Kobayashi:2025ldi,Jiang:2025psz}. 
In the case of $\mathbb{Z}_2$ gauging of a $\mathbb{Z}_N$ symmetry, the resulting Majorana neutrino mass textures are severely restricted, as the Weinberg operator necessarily generates diagonal mass terms. On the other hand, non-invertible symmetries with $\mathbb{Z}_3$ gauging offer a new perspective, allowing a broader class of neutrino mass textures~\cite{Dong:2025jra}. A key difference is that $\mathbb{Z}_3$ gauging admits complex representations, in contrast to the real representations characteristic of $\mathbb{Z}_2$ gauging.

We have studied neutrino mass textures derived from $\mathbb{Z}_3$ gauging of $\mathbb{Z}_N$ symmetries. Two scenarios were considered, in which neutrino masses originate from the Weinberg operator and from the type-I seesaw mechanism, respectively. In the Weinberg-operator case, the neutrino mass texture $\textbf{A}_{1,2}$, $\textbf{B}_{3,4}$ and $\textbf{C}$ can be realized through $\mathbb{Z}_3$ gauging of $\mathbb{Z}_{13}$, while all seven phenomenologically allowed two-zero textures can be successfully obtained in the $\mathbb{Z}_3$ gauging of $\mathbb{Z}_{19}$ symmetry. 
In particular, for $N = 19$, supersymmetry is not required to obtain both diagonal charged-lepton mass matrices and two-zero texture neutrino mass patterns.
When neutrino masses arise from the type-I seesaw mechanism, the texture zeros in $M_\nu$ originate from zero entries in the Dirac neutrino mass matrix through the seesaw formula. Owing to the intrinsic structure of the $\mathbb{Z}_3$ gauging of $\mathbb{Z}_N$ symmetry, only the texture $\mathbf{C}$ can be realized for $N \neq 7$. Furthermore, the Dirac CP-violating phase is restricted to a narrow region, as shown in figure~\ref{fig:chi2plot}.

Furthermore, we investigate the $\mathbb{Z}_n$ gauging of $\mathbb{Z}_N$ for general $n$. Through a detailed study of the automorphism group of the cyclic group $\mathbb{Z}_N$, we generalize the previously discussed $\mathbb{Z}_3$ gauging to a generic $\mathbb{Z}_n$ gauging of $\mathbb{Z}_N$. The richer structure of the automorphism group elucidates the symmetry underlying the corresponding fusion rules. Moreover, by forming tensor products of distinct fusion algebras, one can generate new fusion rules, which frequently admit an interpretation in terms of a larger $\mathbb{Z}_n$ gauging of $\mathbb{Z}_N$. This framework allows for a systematic combination of flavor structures associated with the novel fusion rules.

The interplay between non-invertible symmetries and anomalies deserves closer scrutiny. 
In models where non-invertible symmetries descend from $\mathbb{Z}_N \rtimes \mathbb{Z}_3$, anomaly constraints are expected to be inherited from the parent group, providing a concrete setting to study anomaly matching and possible obstructions.
Another promising direction lies in the dynamical realization and breaking of non-invertible symmetries. Non-invertible symmetries are known to exhibit novel breaking patterns that differ sharply from those of ordinary global symmetries. 
Investigating how these patterns manifest in models with non-invertible $\mathbb{Z}_N$ symmetry origin, and whether partial breaking can interpolate between invertible and non-invertible regimes, may offer new insights into the origin of fermion mass hierarchies.
We leave these issues for future work.

\section*{Acknowledgements}

This work is supported by the National Natural Science Foundation of China under Grant Nos. 12375104, 12547106 and Guizhou Provincial Major Scientific and Technological
Program XKBF (2025)010.

\clearpage

\begin{appendix}

\section{The allowed regions of $\theta_{23}$ and $\delta_\text{CP}$ for the two-zero textures of the Majorana neutrino mass matrix  \label{sec:phenomenon}}

As shown in Eq.~\eqref{eq:ABC-2zero}, the phenomenologically viable light neutrino mass matrices with two independent texture zeros can be classified into seven distinct patterns, namely $\textbf{A}_{1,2}$, $\textbf{B}_{1,2,3,4}$, and $\textbf{C}$, in the basis where the charged-lepton mass matrix is diagonal. The vanishing matrix elements impose strong constraints on the neutrino mass and mixing parameters~\cite{Frampton:2002yf,Xing:2002ta,Fritzsch:2011qv}. In the following, we present the predictions for the atmospheric mixing angle $\theta_{23}$ and the Dirac CP-violating phase $\delta_{\text{CP}}$ for each texture.

\begin{figure}[t!]
\centering
\includegraphics[scale=1.1]{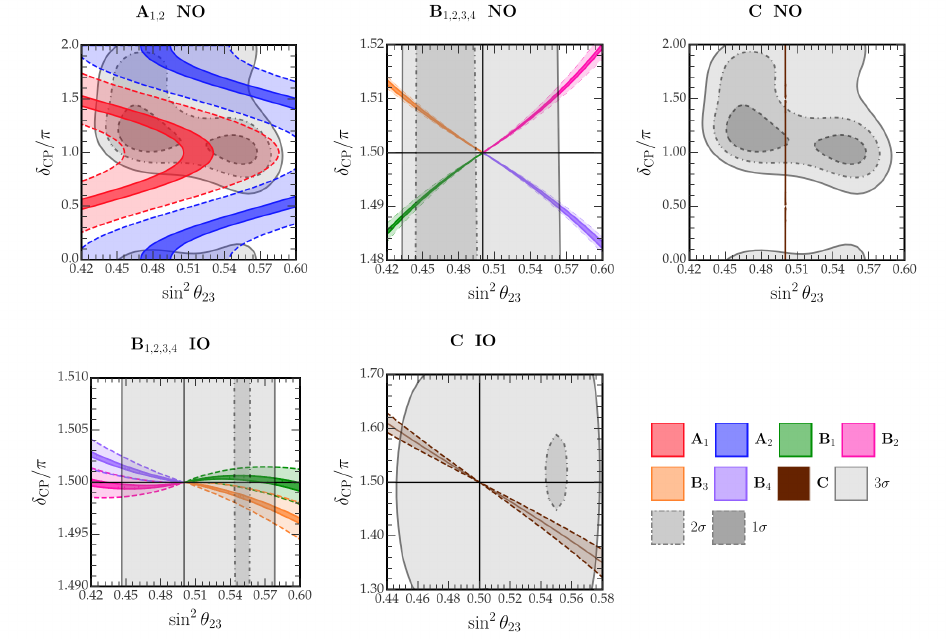}
\caption{The allowed regions of $\delta_{\text{CP}}$ and $\sin^2\theta_{23}$ extracted from Eq.~\eqref{eq:rconstraint} for different two-zero textures of neutrino mass matrix, $\delta m^2/|\Delta m^2|$ freely varies in its $3\sigma$ region. The gray regions denote the experimentally favored $1\sigma$, $2\sigma$ and $1\sigma$ regions of  $\delta_{\text{CP}}$ and $\sin^2\theta_{23}$~\cite{Esteban:2024eli}. The darker regions are obtained by fixing $\theta_{13}$ and $\theta_{12}$ at their best-fit values, while the lighter regions are obtained by varying $\theta_{13}$ and $\theta_{12}$ within their $3\sigma$ ranges~\cite{Esteban:2024eli}. 
}
\label{fig:chi2plot}
\end{figure}

Under the assumption that light neutrinos are Majorana particles, the Lagrangian for the charged lepton and neutrino masses is given by Eq.~\eqref{eq:Lag-charged-lepton-neutrino}. After the electroweak symmetry breaking, the charged-lepton mass matrix $M_E$ and Majorana neutrino mass matrix $M_{\nu}$ can be diagonalized as follow, 
\begin{equation}
U_{\ell L}^\dagger M_E 
U_{\ell R}
\equiv\begin{pmatrix}
        m_e & 0 & 0 \\
        0 & m_\mu & 0 \\
        0 & 0 & m_\tau
    \end{pmatrix} \,,\,\,
    U_{\nu L}^\dagger M_\nu U_{\nu L}^*\equiv\begin{pmatrix}
        m_1 & 0 & 0 \\
        0 & m_2 & 0 \\
        0 & 0 & m_3
    \end{pmatrix} \,,
\end{equation}
where $m_{e,\mu,\tau}$ and $m_{1,2,3}$ stand for the mass eigenvalues of charged leptons and neutrinos respectively, $U_{\ell L}$, $U_{\ell R}$ and $ U_{\nu L}$ are three dimensional unitary matrices. In the mass eigenstate basis, the lepton mixing matrix 
$U_\text{PMNS}$ appears in the flavor-changing charged-current weak interactions of leptons, and is given by
\begin{equation}
U_\text{PMNS}=U_{\ell L}^\dagger U_{\nu L} \,,
\end{equation}
where $U_\text{PMNS}$ is also known as the Pontecorvo–Maki–Nakagawa–Sakata (PMNS) matrix. The PMNS matrix can be parameterized as~\cite{ParticleDataGroup:2024cfk}
\begin{equation}
U_\text{PMNS}=UP
 \,,
\end{equation}
with 
\begin{equation}
    U\equiv\begin{pmatrix}
c_{12}c_{13} & s_{12}c_{13} & s_{13}e^{-i\delta_\text{CP}} \\
-s_{12}c_{23}-c_{12}s_{23}s_{13}e^{i\delta_\text{CP}} & c_{12}c_{23}-s_{12}s_{23}s_{13}e^{i\delta_\text{CP}} & s_{23}c_{13} \\
s_{12}s_{23}-c_{12}c_{23}s_{13}e^{i\delta_\text{CP}} & -c_{12}s_{23}-s_{12}c_{23}s_{13}e^{i\delta_\text{CP}} & c_{23}c_{13}
\end{pmatrix} \,,\quad
P\equiv
\begin{pmatrix}
 1 & 0 & 0 \\
 0 & e^{i\alpha_{21}/2} & 0 \\
 0 & 0 & e^{i\alpha_{31}/2}
\end{pmatrix} \,,
\end{equation}
where $c_{ij}=\cos\theta_{ij}$, $s_{ij}=\sin\theta_{ij}$,  $\theta_{ij}\in[0, \pi/2)$ are the three lepton mixing angles, $\delta_\text{CP}\in[0, 2\pi)$ is the Dirac CP violation phase, and $\alpha_{21}$, $\alpha_{31}$ are two Majorana CP violation phases. In the charged-lepton diagonal basis, $U_{\ell L}$ is a diagonal unitary matrix and $U_\text{PMNS}=U_{\nu L}$. As a result, the light neutrino mass matrix $M_{\nu}$ can be expressed in terms of light neutrino masses and lepton mixing matrix as  
\begin{eqnarray}
M_\nu
&=&UP\,\text{diag}(m_1, m_2, m_3)\,P^TU^T \,, \label{eq:Mnu-U-mhat}
\end{eqnarray}
which gives
\begin{equation}
\label{eq:Mnu-ab}(M_\nu)_{ab}=\sum_{i=1,2,3}m_ie^{i\alpha_{i1}}U_{ai}U_{bi} \,,
\end{equation}
where $\alpha_{i1}=0$ when $i=1$. The two zero entries in the textures $\textbf{A}_{1,2}$, $\textbf{B}_{1,2,3,4}$, $\textbf{C}$ allow to correlate neutrino masses with mixing parameters. For example, if $(M_\nu)_{ab}=(M_\nu)_{cd}=0$ with $ab\neq cd$, from Eq.~\eqref{eq:Mnu-ab}, we can obtain
\begin{eqnarray}
    &&m_1U_{a1}U_{b1}+m_2e^{i\alpha_{21}}U_{a2}U_{b2}+m_3e^{i\alpha_{31}}U_{a3}U_{b3}=0 \,,\nonumber\\
    &&m_1U_{c1}U_{d1}+m_2e^{i\alpha_{21}}U_{c2}U_{d2}+m_3e^{i\alpha_{31}}U_{c3}U_{d3}=0 \,,
\end{eqnarray}
which lead to 
\begin{eqnarray}
\frac{m_1}{m_3}e^{-i\alpha_{31}}=\frac{U_{a3}U_{b3}U_{c2}U_{d2}-U_{a2}U_{b2}U_{c3}U_{d3}}{U_{a2}U_{b2}U_{c1}U_{d1}-U_{a1}U_{b1}U_{c2}U_{d2}} \,,\nonumber\\
\frac{m_2}{m_3}e^{i(\alpha_{21}-\alpha_{31})}=\frac{U_{a1}U_{b1}U_{c3}U_{d3}-U_{a3}U_{b3}U_{c1}U_{d1}}{U_{a2}U_{b2}U_{c1}U_{d1}-U_{a1}U_{b1}U_{c2}U_{d2}} \,.
\end{eqnarray}
It follows that the neutrino mass ratios can be determined from the neutrino mixing angles and the Dirac CP-violating phase as follows:
\begin{eqnarray}
r_{13}\equiv\frac{m_1}{m_3}=\left|\frac{U_{a3}U_{b3}U_{c2}U_{d2}-U_{a2}U_{b2}U_{c3}U_{d3}}{U_{a2}U_{b2}U_{c1}U_{d1}-U_{a1}U_{b1}U_{c2}U_{d2}}\right| \,,\nonumber\\
r_{23}\equiv\frac{m_2}{m_3}=\left|\frac{U_{a1}U_{b1}U_{c3}U_{d3}-U_{a3}U_{b3}U_{c1}U_{d1}}{U_{a2}U_{b2}U_{c1}U_{d1}-U_{a1}U_{b1}U_{c2}U_{d2}}\right| \,,\label{eq:massratio}
\end{eqnarray}
and the Majorana CP-violating phases are
\begin{eqnarray}
    &&\alpha_{31}=-\text{arg}\left[\frac{U_{a3}U_{b3}U_{c2}U_{d2}-U_{a2}U_{b2}U_{c3}U_{d3}}{U_{a2}U_{b2}U_{c1}U_{d1}-U_{a1}U_{b1}U_{c2}U_{d2}}\right] \,,\nonumber\\
    &&\alpha_{21}=\alpha_{31}+\text{arg}\left[\frac{U_{a1}U_{b1}U_{c3}U_{d3}-U_{a3}U_{b3}U_{c1}U_{d1}}{U_{a2}U_{b2}U_{c1}U_{d1}-U_{a1}U_{b1}U_{c2}U_{d2}}\right] \,.\label{eq:majorcpviophase}
\end{eqnarray}
From Eq.~\eqref{eq:Mnu-U-mhat}, we see that the Majorana neutrino mass matrix depends on nine physical parameters including three neutrino masses, three mixing angles and three CP violation phases. The two vanishing entries in neutrino mass matrix give rise to four constraints, namely Eqs.~\eqref{eq:massratio} and \eqref{eq:majorcpviophase}, leaving five independent parameters, which exactly match the five experimental observables such as three mixing angles and two neutrino mass-squared differences. Therefore, we can use the experimental data to predict the remaining four parameters, in particular the Dirac CP violating phase $\delta_{\text{CP}}$. It is well-known that the neutrino oscillation experiments can measure two neutrino mass squared differences $\delta m^2$ and $\Delta m^2$,
\begin{equation}
\delta m^2\equiv m_2^2-m_1^2 \,,\quad \Delta m^2\equiv m_3^2-\frac{1}{2}(m_2^2+m_1^2) \,,\label{eq:neumasssqdif}\,.
\end{equation}
In combination with Eq.~\eqref{eq:massratio}, the texture-zero conditions establish relations between the neutrino masses and the mixing parameters, which can be written as 
\begin{equation}
    m_3=\sqrt\frac{\delta m^2}{r_{23}^2-r_{13}^2} \,,\quad m_2=m_3r_{23} \,,\quad m_1=m_3r_{13} \,,
\end{equation}
and
\begin{equation}
\frac{\delta m^2}{|\Delta m^2|}=\frac{2(r_{23}^2-r_{13}^2)}{|2-r_{23}^2-r_{13}^2|} \,, \label{eq:rconstraint}
\end{equation}
which correlates the neutrino mass-squared differences with the lepton mixing angles as well as the Dirac CP-violating phase $\delta_{\text{CP}}$, whose value has not yet been experimentally fixed. Both ratios $r_{13}$ and $r_{23}$ are invariant under the transformation  $U\rightarrow U^*$ (equivalently $\delta_\text{CP}\rightarrow-\delta_\text{CP}$), therefore the relation of Eq.~\eqref{eq:rconstraint} is insensitive to the sign of $\delta_\text{CP}$. Given that $\delta m^2$ and $\Delta m^2$, $\theta_{13}$ and $\theta_{12}$ have been measured with high precision~\cite{Esteban:2024eli},  
one can extract the allowed regions for $\theta_{23}$ and $\delta_{\text{CP}}$ from Eq.~\eqref{eq:rconstraint}. The viable parameter space of $\theta_{23}$ and $\delta_{\text{CP}}$ is displayed in figure~\ref{fig:chi2plot} for different textures with two zero entries. 

For the textures $\textbf{A}_{1}$ and $\textbf{A}_{2}$, only normal ordering (NO) neutrino mass spectrum can be accommodated, the atmospheric angle $\theta_{23}$ and Dirac CP phase $\delta_\text{CP}$ can vary in large regions, if the uncertainties of $\theta_{12}$ and $\theta_{13}$ are taken into account. For the two-zero textures $\textbf{B}_{1,2,3,4}$, the neutrino mass spectrum can be either NO or inverted ordering (IO), and Eq.~\eqref{eq:rconstraint} constrains the Dirac CP violation phase $\delta_{\text{CP}}$ to be around $1.5\pi$ or $0.5\pi$. In order to avoid large blank regions in the plots, we show only the region around $\delta_{\text{CP}}=1.5\pi$ and omit the region around $\delta_{\text{CP}}=0.5\pi$ in the panels for $\textbf{B}_{1,2,3,4}$ in figure~\ref{fig:chi2plot}. Moreover, $\theta_{23}$ and $\delta_{\text{CP}}$ display a significant correlation. In the case of NO, these textures lead to definite predictions for the signs of $\cos{2\theta_{23}}$ and $\cos{\delta_\text{CP}}$ such as,
\begin{eqnarray}
\textbf{B}_1&:& \quad \theta_{23}<45^\circ \,,\,\, \delta_\text{CP}<1.5 \, \pi\,, \nonumber\\
\textbf{B}_2&:& \quad \theta_{23}>45^\circ \,,\,\, \delta_\text{CP}>1.5 \, \pi\,, \nonumber\\
\textbf{B}_3&:& \quad \theta_{23}<45^\circ \,,\,\, \delta_\text{CP}>1.5 \, \pi\,, \nonumber\\
\textbf{B}_4&:& \quad \theta_{23}>45^\circ \,,\,\, \delta_\text{CP}<1.5 \, \pi \,.
\end{eqnarray}
In the case of IO neutrino mass spectrum, these textures impose the following constraints: $\textbf{B}_1$ and $\textbf{B}_3$ favor the higher octant ($\theta_{23}>45^\circ$), whereas $\textbf{B}_2$ and $\textbf{B}_4$ prefer the lower octant ($\theta_{23} < 45^\circ$); in addition, $\textbf{B}_3$ predicts $\delta_{\text{CP}} < 1.5\,\pi$ and $\textbf{B}_4$ predicts $\delta_{\text{CP}} > 1.5\,\pi$.

As for the texture $\textbf{C}$, both NO and IO neutrino mass can be compatible with the experimental data. In the NO case, expanding the right-hand side of Eq.~\eqref{eq:rconstraint} around $\theta_{23}=45^\circ$, we obtain:
\begin{equation}
\frac{\delta m^2}{|\Delta m^2|}\approx-\frac{\cos^4{\theta_{13}}\sec{\delta_\text{CP}}}{\sin^3{\theta_{13}}\sin{\theta_{12}}\cos{\theta_{12}}}(\theta_{23}-45^\circ)\,.\label{eq:texCleadingorder}
\end{equation} 
Using the results of the recent global analysis of neutrino oscillation data~\cite{Esteban:2024eli}, we have $\delta m^2/|\Delta m^2| \approx 0.03$ and $\cos^4{\theta_{13}}\sec{\delta_\text{CP}}/(\sin^3{\theta_{13}}\sin{\theta_{12}}\cos{\theta_{12}})\geq538$, which implies that $|\theta_{23} - 45^\circ|$ is of order $\mathcal{O}(10^{-4})$. Consequently, the atmospheric mixing angle $\theta_{23}$ is extremely close to $45^\circ$, while $\delta_{\text{CP}}$ is essentially unconstrained except for $\delta_{\text{CP}} \neq 0.5\pi\,, 1.5\pi$. As a result, the allowed parameter region appears as a vertical line in the panel $\textbf{C} \,\,\, \textbf{NO}$. It should be emphasized that the exact points $(\theta_{23}\,,\delta_{\text{CP}}) = (45^\circ\,, 1.5\pi)$ and $(45^\circ\,, 0.5\pi)$ cannot be realized in this scenario, since the three neutrino masses become degenerate at these parameter values. 
Finally, in the IO case for the texture $\textbf{C}$, a strong correlation between $\theta_{23}$ and $\delta_{\text{CP}}$ emerges. In particular, the atmospheric angle lying in the higher (lower) octant, i.e. $\theta_{23}>(<)\,45^\circ$, implies $\delta_{\text{CP}}<(>)\,1.5\,\pi$. This correlation can also be inferred from Eq.~\eqref{eq:texCleadingorder}.

\section{$\mathbb{Z}_3$ gauging of $\mathbb{Z}_N$ symmetries for $N=7$ and $9$\label{sec:Z3_gauging_of_Z7_Z9}}

\subsection{$N = 7$}

When $N=7$, $m=2$ or $m=4$, there are three equivalence classes: $[0]$, $[1]$ and $[3]$, 
\begin{eqnarray}\label{eq:Z7_class}
[0] = \{ e \}\,,~~~[1] = \{a, a^2, a^4\}\,,~~~ [3] = \{ a^3, a^5, a^6 \} \,,
\end{eqnarray}
and they obey the fusion rules
\begin{eqnarray}
\nonumber&&\qquad\qquad   [0] \times [k] = [k]\,,~~ [1]\times [1] = [1] + 2[3]\,,~~ \\
\label{eq:fussion-rule-N7} && [3]\times [3] = [3] + 2 [1]\,,~~ [1]\times [3] = [3] \times[1] = 3[0] + [1] + [3] \,,
\end{eqnarray}
with $k=0, 1, 3$. It indicates that there exists the $\mathbb{Z}_2$ permutation symmetry associated with $[1]\leftrightarrow [3]$. 
This $\mathbb{Z}_2$ corresponds to the outer automorphism exchanging $a$ and $a^{-1}$. The equivalence classes in Eq.~\eqref{eq:Z7_class} correspond to the conjugacy classes of $\mathbb{Z}_7 \rtimes \mathbb{Z}_3 \cong T_7$ in the absence of the element $b$.

As discussed in section~\ref{sec:neu_mass}, if two or three generations of $L_i$ correspond to the same class, the neutrino mass matrix can be of the patterns in Eq.~\eqref{eq:texture-L1L2same} and the desired two-zero textures can not be produced if light neutrino mass is described by Weinberg operator. If the left-handed leptons are assigned to distinct equivalence classes, the only allowed assignment is $L_i \sim ([0], [1], [3])$ up to permutations. The Higgs field can correspond to either $[0]$ or $[1]$. The textures are
\begin{eqnarray}\label{eq:Weinberg_N7}
H \sim [0]:~ M_\nu = \begin{pmatrix}
\times ~&~ 0 ~&~ 0  \\
0 ~&~ 0 ~&~ \times  \\
0 ~&~ \times ~&~ 0  
\end{pmatrix} \,,~~~~
H\sim [1]:~ M_\nu = \begin{pmatrix}
0 ~&~ \times ~&~ \times  \\
\times ~&~ \times ~&~ \times  \\
\times ~&~ \times ~&~ \times  
\end{pmatrix} \,.
\end{eqnarray}
If neutrino mass is generated by the type-I seesaw mechanism, the textures $\textbf{A}_{1,2}$, $\textbf{B}_{1,2,3,4}$ and $\textbf{C}$ cannot be obtained at $N=7$, as shown in section~\ref{sec:seesaw}. 

Moreover, when considering the $\mathbb{Z}_3$ gauging of $\mathbb{Z}_N$ with $N=14$, the corresponding fusion rules are given by the product of $\mathbb{Z}_2$ group fusion rule and the the fusion rules from $\mathbb{Z}_3$ gauging of $\mathbb{Z}_7$. Nevertheless, the presence of the extra $\mathbb{Z}_2$ symmetry does not facilitate the appearance of texture zeros. As a consequence, because the $\mathbb{Z}_3$ gauging of $\mathbb{Z}_{14}$ inherits the algebraic structure of the fusion rules in $\mathbb{Z}_3$ gauging of $\mathbb{Z}_{7}$, the experimentally viable neutrino mass with two texture zeros still cannot be realized.

\subsection{$N = 9$}

We study the case of $N=9, m=4$, where there are five classes $[0]$, $[1]$, $[2]$, $[3]$ and $[6]$. The same set of equivalence classes are reached for $m=7$. Using Eq.~\eqref{eq:EQ-classes}, we can obtain the equivalence classes as follow,
\begin{eqnarray}
[0] = \{e\} \,,~~~[3] = \{a^3\} \,,~~~[6] = \{a^6\}\,,~~~ [1] = \{a, a^4, a^7\}\,,~~~ [2]= \{a^2, a^5, a^8 \}\,.
\end{eqnarray}
From Eq.~\eqref{eq:fussion-rules-gener}, we know the corresponding fusion rules are given by
\begin{eqnarray}
\nonumber &&[0]\times [n]=[n]\,,~ [3] \times [3] = [6] \,,~ [6] \times [6] = [3]\,,~[3] \times [6] = [0]\,,~ 
[3] \times [1] = [6] \times [1] = [1]\,, \\
&&[3] \times [2] = [6] \times [2] = [2] \,,~ [1] \times [1] = 3[2] \,,~ [2] \times [2] = 3[1] \,,~ [1] \times [2] = 3[0] + 3[3] + 3[6]  \,.~~~~~~
\end{eqnarray}
There also exists a $\mathbb{Z}_2 \times \mathbb{Z}_2$ symmetry associated with $[1]\leftrightarrow [2]$ and $[3]\leftrightarrow [6]$. It should be noted that $[0]$, $[3]$, $[6]$ formulate a closed $\mathbb{Z}_3$ group fusion rule. We assume that the three generations of left-handed lepton fields are assigned to different equivalent classes. If the light neutrino masses are described by the Weinberg operator, the possible textures of the light neutrino mass matrix is listed in table~\ref{tab:Z9_Wein}.

From table~\ref{tab:Z9_Wein}, we observe that for all assignments, each neutrino mass matrix contains more than three texture zeros.
As discussed in Sec~\ref{sec:seesaw}, for a fixed assignment of the left-handed lepton fields, the neutrino mass matrix generated via seesaw mechanism generically contains more zeros than the corresponding mass matrix arising from Weinberg operator.
Consequently, the textures $\textbf{A}_{1,2}$, $\textbf{B}_{1,2,3,4}$ and $\textbf{C}$ cannot be realized through the seesaw mechanism for $N=9$.

\begin{table}[h!]
\centering
\resizebox{0.65\textwidth}{!}{
\begin{tabular}{|c|c|c||c|c|c|}		
	\hline\hline
$M_\nu$	& $H\sim[0]$ & $H\sim[1]$ & $H\sim[3]$ \\ \hline		
	$\ell_L\sim([0],[3],[6])$ &
	$\begin{pmatrix}
		\times & 0 & 0\\
		0 & 0 & \times\\
		0 & \times & 0
	\end{pmatrix}$ &
	$\begin{pmatrix}
		0 & 0 & 0\\
		0 & 0 & 0\\
		0 & 0 & 0
	\end{pmatrix}$ &
	$\begin{pmatrix}
		0 & \times & 0\\
		\times & 0 & 0\\
		0 & 0 & \times
	\end{pmatrix}$ \\
	\hline
	$\ell_L\sim([0],[3],[1])$ &
	$\begin{pmatrix}
		\times & 0 & 0\\
		0 & 0 & 0\\
		0 & 0 & 0
	\end{pmatrix}$ &
	$\begin{pmatrix}
		0 & 0 & \times\\
		0 & 0 & \times\\
		\times & \times & 0
	\end{pmatrix}$ &
	$\begin{pmatrix}
		0 & \times & 0\\
		\times & 0 & 0\\
		0 & 0 & 0
	\end{pmatrix}$ \\
	\hline
	$\ell_L\sim([0],[3],[2])$ &
	$\begin{pmatrix}
		\times & 0 & 0\\
		0 & 0 & 0\\
		0 & 0 & 0
	\end{pmatrix}$ &
	$\begin{pmatrix}
		0 & 0 & 0\\
		0 & 0 & 0\\
		0 & 0 & \times
	\end{pmatrix}$ &
	$\begin{pmatrix}
		0 & \times & 0\\
		\times & 0 & 0\\
		0 & 0 & 0
	\end{pmatrix}$ \\
	\hline
	$\ell_L\sim([0],[6],[1])$  &
	$\begin{pmatrix}
		\times & 0 & 0\\
		0 & 0 & 0\\
		0 & 0 & 0
	\end{pmatrix}$ &
	$\begin{pmatrix}
		0 & 0 & \times\\
		0 & 0 & \times\\
		\times & \times & 0
	\end{pmatrix}$ &
	$\begin{pmatrix}
		0 & 0 & 0\\
		0 & \times & 0\\
		0 & 0 & 0
	\end{pmatrix}$ \\
	\hline
	$\ell_L\sim([0],[6],[2])$  &
	$\begin{pmatrix}
		\times & 0 & 0\\
		0 & 0 & 0\\
		0 & 0 & 0
	\end{pmatrix}$ &
	$\begin{pmatrix}
		0 & 0 & 0\\
		0 & 0 & 0\\
		0 & 0 & \times
	\end{pmatrix}$ &
	$\begin{pmatrix}
		0 & 0 & 0\\
		0 & \times & 0\\
		0 & 0 & 0
	\end{pmatrix}$ \\
	\hline
	$\ell_L\sim([3],[6],[1])$ &
	$\begin{pmatrix}
		0 & \times & 0\\
		\times & 0 & 0\\
		0 & 0 & 0
	\end{pmatrix}$ &
	$\begin{pmatrix}
		0 & 0 & \times\\
		0 & 0 & \times\\
		\times & \times & 0
	\end{pmatrix}$ &
	$\begin{pmatrix}
		0 & 0 & 0\\
		0 & \times & 0\\
		0 & 0 & 0
	\end{pmatrix}$ \\
	\hline
	$\ell_L\sim([3],[6],[2])$  &
	$\begin{pmatrix}
		0 & \times & 0\\
		\times & 0 & 0\\
		0 & 0 & 0
	\end{pmatrix}$ &
	$\begin{pmatrix}
		0 & 0 & 0\\
		0 & 0 & 0\\
		0 & 0 & \times
	\end{pmatrix}$ &
	$\begin{pmatrix}
		0 & 0 & 0\\
		0 & \times & 0\\
		0 & 0 & 0
	\end{pmatrix}$ \\
	\hline
	$\ell_L\sim([0],[1],[2])$ &
	$\begin{pmatrix}
		\times & 0 & 0\\
		0 & 0 & \times\\
		0 & \times & 0
	\end{pmatrix}$ &
	$\begin{pmatrix}
		0 & \times & 0\\
		\times & 0 & 0\\
		0 & 0 & \times
	\end{pmatrix}$ &
	$\begin{pmatrix}
		0 & 0 & 0\\
		0 & 0 & \times\\
		0 & \times & 0
	\end{pmatrix}$ \\
	\hline
	$\ell_L\sim([3],[1],[2])$  &
	$\begin{pmatrix}
		0 & 0 & 0\\
		0 & 0 & \times\\
		0 & \times & 0
	\end{pmatrix}$ &
	$\begin{pmatrix}
		0 & \times & 0\\
		\times & 0 & 0\\
		0 & 0 & \times
	\end{pmatrix}$ &
	$\begin{pmatrix}
		0 & 0 & 0\\
		0 & 0 & \times\\
		0 & \times & 0
	\end{pmatrix}$ \\
	\hline
	$\ell_L\sim([6],[1],[2])$  &
	$\begin{pmatrix}
		0 & 0 & 0\\
		0 & 0 & \times\\
		0 & \times & 0
	\end{pmatrix}$ &
	$\begin{pmatrix}
		0 & \times & 0\\
		\times & 0 & 0\\
		0 & 0 & \times
	\end{pmatrix}$ &
	$\begin{pmatrix}
		\times & 0 & 0\\
		0 & 0 & \times\\
		0 & \times & 0
	\end{pmatrix}$ \\
	\hline\hline
\end{tabular}}
\caption{\label{tab:Z9_Wein}The neutrino mass matrix $M_\nu$ from $\mathbb{Z}_3$ gauging of $\mathbb{Z}_9$ symmetry, where the neutrino mass is described by the Weinberg operator.  }
\end{table}

\section{More general $\mathbb{Z}_2$ gauging of $\mathbb{Z}_N$ symmetries \label{sec:Z2_gauging_of_ZN}}

In the $\mathbb{Z}_2$ gauging of $\mathbb{Z}_N$ symmetry, usually the following automorphism is considered 
\begin{eqnarray}\label{eq:usual_Z2}
b^2 = e\,,~~~ b^{-1} a b = a^{-1} \,.
\end{eqnarray}
However, this is not unique. The most general automorphism for the $\mathbb{Z}_2$ gauging of $\mathbb{Z}_N$ is given by
\begin{eqnarray}
b^2 = e \,,~~~ b^{-1} a b = a^{m} \,.
\end{eqnarray}
with 
\begin{eqnarray}\label{eq:cond-2GaugeN}
m^2 - 1 \equiv 0 ~({\rm mod}~N)  \,.
\end{eqnarray}
When $N=p^\alpha$ with prime number $p\geq 3$ or $p=2$, $\alpha = 2$, Eq.~\eqref{eq:AuT-Zpalpha} shows ${\rm Aut}(\mathbb{Z}_{N}) \cong \mathbb{Z}_{(p-1)p^{\alpha-1}}$. In this case, the automorphism group ${\rm Aut}(\mathbb{Z}_{N})$ contains a unique $\mathbb{Z}_2$ subgroup, and Eq.~\eqref{eq:cond-2GaugeN} admits only the solution $m=-1$.
In contrast, for $N=2^{\alpha}$ with $\alpha\geq 3$, the automorphism group is not a cyclic group and instead satisfies ${\rm Aut}(\mathbb{Z}_{2^\alpha}) \cong \mathbb{Z}_2 \times \mathbb{Z}_{2^{\alpha-2}}$ as stated in Eq.~\eqref{eq:Aut_2alpha}. Then there exist three inequivalent $\mathbb{Z}_2$ subgroups, corresponding to three distinct values of $m$, namely
\begin{eqnarray}
m = -1\,, \qquad m = 2^{\alpha-1}-1\,, \qquad m = 2^{\alpha-1}+1\,.
\end{eqnarray}
The choice $m=-1$ reproduces the automorphism given in Eq.~\eqref{eq:usual_Z2}, and the group generated by $a$ and $b$ is isomorphic to the dihedral group $D_N$. For the second solution $m = 2^{\alpha-1}-1$, the resulting group generated by $a$ and $b$ is isomorphism to the quasi-dihedral group $QD_{2N}$~\cite{Ishimori:2010au}.
As an explicit illustration, we examine the case $N=8$ with $\alpha=3$.
For $m=-1$, we recover the usual $\mathbb{Z}_2$ gauging, in which $a^k$ and $a^{-k}$ belong to the same equivalence class,
\begin{eqnarray}
[0]=\{e\}\,,~~~ [1] = \{a,a^7\}\,,~~~ [2] = \{a^2, a^6\}\,,~~~ [3] = \{a^3, a^5\}\,,~~~ [4] = \{a^4\} \,.
\end{eqnarray}
The fusion rules are given by
\begin{eqnarray}
\nonumber && [1]\times [1] = [3] \times [3] = 2[0] + [2]\,,~ 
[1] \times [3] = [2] + 2[4] \,,~ 
[1] \times [2] = [3] \times [2] = [1] + [3] \,,~~~ \\
&& [2] \times [2] = 2[0] + 2[4] \,,~ 
[1] \times [4] = [3] \,,~
[3] \times [4] = [1] \,,~
[2] \times [4] = [2] \,,~
[4] \times [4] = [0] \,.
\end{eqnarray}
The above fusion rules possess a $\mathbb{Z}_2$ symmetry under the exchange $[1] \leftrightarrow [3]$.

If we take $m = 2^{\alpha - 1} - 1 =3$, the equivalence classes are
\begin{eqnarray}
[0]=\{e\}\,,~~~ [1] = \{a, a^3\}\,,~~~ [2] = \{a^2, a^6\}\,,~~~ [4] = \{a^4\}\,,~~~ [5] = \{a^5, a^7\} \,.
\end{eqnarray}
which satisfy the following fusion rules, \begin{eqnarray}
\nonumber && [1] \times [1] = [5] \times [5] = [2] + 2[4] \,,~
[1] \times [5] = 2[0] + [2] \,,~
[1] \times [2] = [5] \times [2] = [1] + [5] \,,~~~  \\
&& [2] \times [2] = 2[0] + [4] \,,~
[1] \times [4] = [5] \,,~
[5] \times [4] = [1] \,,~
[2] \times [4] = [2] \,,~
[4] \times [4] = [0] \,.
\end{eqnarray}
These fusion rules further exhibit a $\mathbb{Z}_2$
symmetry under the exchange $[1] \leftrightarrow [5]$.

In the case of $m = 2^{\alpha - 1} + 1 =5$, the equivalence classes are
\begin{eqnarray}
[0]=\{e\}\,,~~ [1] = \{a,a^5\}\,,~~ [2] = \{a^2\}\,,~~ [3] = \{a^3, a^7\}\,,~~ [4] = \{a^4\}\,,~~ [6] = \{a^6\} \,.~~~
\end{eqnarray}
The associated fusion rules read as 
\begin{eqnarray}
\nonumber && [1] \times [1] = [3] \times [3] = 2[2] + 2[6] \,,~~~
[1] \times [3] = 2[0] + 2[4] \,,~~~
[1] \times [4] = [1] \,,~~~ 
[3] \times [4] = [3] \,,~~~ \\
\nonumber &&[1] \times [2] = [1] \times [6] = [3] \,,~~~ 
[3] \times [2] = [3] \times [6] = [1] \,,~~~   
[2] \times [2] = [6] \times [6] = [4] \,,~~~ \\
&& [2] \times [6] = [4] \times [4] = [0] \,,~~~
[2] \times [4] = [6] \,,~~~
[4] \times [6] = [2]  \,,
\end{eqnarray}
which exhibit a $\mathbb{Z}_2 \times \mathbb{Z}_2$ symmetry generated by the interchanges $[1] \leftrightarrow [3]$ and $[2] \leftrightarrow [6]$. Moreover, the equivalence classes $[0]$, $[2]$, $[4]$, and $[6]$ form a $\mathbb{Z}_4$ subgroup.

\clearpage

\end{appendix}

\providecommand{\href}[2]{#2}\begingroup\raggedright\endgroup

\end{document}